\documentclass[12pt]{article}

\usepackage{graphicx}

\usepackage{subfigure}

\usepackage{epsfig}

\begin{document}

\centerline{\bf Agent-based Models of Financial Markets}

\centerline{E. Samanidou$^1$, E. Zschischang$^2$, D. Stauffer$^3$ , and T. Lux$^4$}
\small{

\noindent

$^1$ Deutsche Bundesbank, Referent Bankgesch\"aftliche Pr\"ufungen, Berliner Allee 40,D-40212 D\"usseldorf

\noindent

$^2$ HSH Nord Bank, Portfolio Mngmt. \& Inv., Martensdamm 6, D-24103 Kiel

\noindent

$^3$ Institute for Theoretical Physics, Cologne University, D-50923 K\"oln}

$^4$ Department of Economics, University of Kiel, Olshausenstrasse 40, D-24118 Kiel

\begin{abstract} \label{abstract}
This review deals with several microscopic (``agent-based'') models of financial markets which have been studied
by economists and physicists over the last decade: Kim-Markowitz, Levy-Levy-Solomon,
Cont-Bouchaud, Solomon-Weisbuch, Lux-Marchesi, Donangelo-Sneppen and Solomon-Levy-Huang.
After an overview of simulation approaches in financial economics, we first give a summary of
the Donangelo-Sneppen model of monetary exchange and compare it with related models in economics
literature. Our selective review then outlines the main ingredients of some influential
early models of multi-agent dynamics in financial markets (Kim-Markowitz, Levy-Levy-Solomon).
 As will be seen, these contributions draw their inspiration from the complex appearance of
 investors' interactions in real-life markets. Their main aim is to reproduce (and, thereby,
 provide possible explanations) for the spectacular bubbles and crashes seen in certain
 historical episodes, but they lack (like almost all the work before 1998 or so) a
 perspective in terms of the universal statistical features of financial time series.
 In fact, awareness of a set of such regularities (power-law tails of the  distribution of
 returns, temporal scaling of volatility) only gradually appeared over the nineties.
 With the more precise description of the formerly relatively vague characteristics (
 e.g. moving from the notion of fat tails to the more concrete one of a power-law with
 index around three), it became clear that financial markets dynamics give rise to some
 kind of universal scaling laws. Showing similarities with scaling laws for other systems
 with many interacting sub-units, an exploration of financial markets as multi-agent systems
 appeared to be a natural consequence. This topic was pursued by quite a number of contributions
 appearing in both the physics and economics literature since the late nineties. From the wealth
  of different flavors of multi-agent models that have appeared by now, we discuss the
  Cont-Bouchaud, Solomon-Levy-Huang and Lux-Marchesi models. Open research questions are
  discussed in our concluding section.

\end{abstract}

\section{Introduction}\label{Intro}

Physicists not only know everything, they also know everything better. This indisputable dogma does not exclude, however, that some economists published work similar to what physicists now celebrate as ``econophysics'', only much earlier, like Nobel laureate Stigler \cite{Stigler} (which was not exactly agent based; nor are
all econophysics models agent-based). Are econophysicists like Christopher Columbus, rediscovering something which others had found earlier, and also getting things somewhat wrong, but nevertheless changing human history? As the team of authors of this survey collects scientists from both disciplines, we do not attempt to give a definite answer to this question, but simply review some influential models by both physicists and economists, to allow a fair comparison.
Long ago, according to \cite{Ingrao}, economists like Walras and Pareto were
inspired by Newtonian mechanics.

Stylized facts is the economist's name for universal properties of markets, independent of whether we look at New York, Tokyo, or Frankfurt, or whether we are concerned with share markets, foreign exchange markets or derivative markets. The following is a collection of those ``stylized facts'' that are now almost universally accepted among economists and physicists:
(i) There is widespread agreement that we cannot predict whether the price tomorrow will go up or down, on the base of past price trends or other current information.
(ii) If today the market had been very volatile, then the probability for observing a large change (positive or negative) tomorrow is also higher than on average (volatility clustering).
(iii) The probability to have a large change in the market, by at least $x\%$, decays with a power law in $1/x$. Fact (iii) has first been discovered by Mandelbrot \cite{Mandelbrot_alt} who proposed the Levy stable model for financial returns. Over the recent years,
the majority opinion (see \cite{Weron,Eisler} for dissent) among researchers in the field has, however, converged to the view that the tails of the cumulative distribution of returns are characterized by a power-law with exponent around 3. The underlying data would, hence,
possess finite variance in contradiction to the Levy stable model. (iv) The $q$-th moments of the distribution of price changes are multifractal, i.e., their exponent is not a linear function of this index $q$ (a rather new
observation). Facts (i) to (iii) can be found in surveys on the
econometrics of financial markets, cf. de Vries \cite{deVries} and Pagan \cite{Pagan}. Fact (iv) has been first partially
documented in Ding {\em et al.} \cite{Ding} and has meanwhile also obtained the status of an
universal feature of all markets in the empirical finance
literature (Lobato and Savin \cite{Lobato}). Similar research on
multiscaling (multifractality), albeit with different analytical tools, was conducted in numerous econophysics papers, starting with Mandelbrot {\em et al.} \cite{Mandelbrot}, Vandewalle and Ausloos \cite{Vandewalle}.

Although research in agent-based models started from a diverse range of intentions (see below), much of the physics-inspired literature considered in this
survey aims at behavioral explanations of the above stylized facts. The more successful ones, in fact, generate even numerically accurate and robust
scaling exponents. It appears worthwhile to point out that with these empirically relevant predictions, the microscopic models meet Friedman's \cite{Friedman} methodological request that
a theory ``(...) is to be judged by its predictive power for the class of phenomena which it is intended to explain''. Despite this conformity with the classical methodological
premise to which most economists pay homage, one might find these models being criticized sometimes because of their lack of ``microfoundations''.
The request of microfoundations in this critique aims at a full-flechted intertemporal optimization as the base of agents' economic activities which mostly
is absent in agent-based models. Hardcore proponents of such a microfoundation would dismiss any theoretical approach that falls short of complete optimization
even if it yields successful preditions. Needless to say that the proponents of agent-based models have a different view and the present authors
would in particular stress the importance of interaction as an alternative facet of microfoundations of macroscopic phenomena in economics. A
dogmatic dismissal of successful theories because of their violation of modeling principles imposed \textit{a priori} would be hard to square with Friedman's
methodological premise to which the same economists would mostly subscribe.

After the pioneering computer simulations of market models from economists like
Stigler \cite{Stigler}, numerous such models were published in the physics literature
since 1992. We concentrate here on those econophysics models which have raised
enough interest to be also investigated by others than the original authors themselves. These include the models of (i) Kim and Markowitz \cite{Kim},
of (ii) Levy, Levy, Solomon \cite{Levy,LevyII,LevyIII,LevyIV,LevyV,LevyVI,LevyVII}, of (iii) Solomon, Levy and Huang \cite{Huang}, of (iv) Cont and Bouchaud \cite{Cont}, of (v)
Solomon and Weisbuch \cite{SolomonIII} (see also \cite{SolW}), and of (vi) Lux and Marchesi \cite{LuxV,LuxVI}, all conceived as models for modern
(financial) markets, as well as the model for ancient barter and
self-organization of monetary exchange by Donangelo and Sneppen \cite{Donangelo} (see also \cite{DonangeloII,BakIII,BakIV}. (Kim and Markowitz are not
econophysicists but used similar methods earlier; Markowitz got a Nobel prize
for portfolio theory but invented his own computer language decades earlier
\cite{Markowitzbook}.)

We start with the latter one since it (in its literal
interpretation) refers to prehistoric times. We neglect the now
(in physics circles) widespread Minority Games, as they arose from
the question when best to visit the El Farol bar in Santa Fe to
avoid overcrowding. The weighted majority of the present
authors prefers to drink experimentally instead of simulating
drinks, and thus we leave these minority games to another review
\cite{Martino}. While according to the late Nobel laureate Friedman an ultimate
aim of models should be to predict the future,
we concentrate here on the easier task of explaining the past; a model which
fails to describe past reality is unlikely to make reliable predictions.

\section{Overview}\label{Overview}

Research applying microscopic simulations in economics
and finance stems from several sources. \footnote{An inspiring source from a field outside economics were the microscopic models of social segregation and
related social phenomena by Schelling (\cite{Schelling}, \cite{SchellingI}). Physicists might recognize its dynamics as a variant of the Ising model.}
First, a number of authors in
the economics ``mainstream'' have resorted to some type of
microscopic simulation in the course of their work on certain
economic problems and models. The framework of Kim and Markowitz \cite{Kim}, explored in detail in section \ref{KimMarkowitz}, may serve as a prominent
example. Like many economists of that time, the authors were
interested in explaining the sudden drop of the U.S. stock market
on 17th October 1987. A widespread explanation for this event was
the automatic overreaction of computer-based ``dynamic hedging'' strategies that had become popular strategies of institutional
investors in the years before. However, models including the
market interactions of many investors following such strategies
are clearly hard to solve in an analytical manner. Therefore, Kim and
Markowitz  decided to investigate the destabilizing potential of
dynamic hedging strategies via Monte Carlo simulations of a
relatively complicated model of price formation in an ``artificial''
financial market (cf. Markowitz \cite{Markowitz}). They were, however, not the first to rely on
simulations of economic processes. During the fifties, the
well-known economist A. W. Phillips -who first recovered the
so-called Phillips curve (i.e., the inverse relationship between
unemployment and inflation rate)- used a hydraulic machine for
simulation of macroeconomic processes (Phillips \cite{Phillips}, see also \cite{Newlyn}). Even earlier, we can find
simulations via electronic circuits published in economics
journals (Morehouse {\em et al.} \cite{Morehouse}).

However, the first simple Monte Carlo simulation of a
financial market appeared in Stigler \cite{Stigler}, who generated trading
orders as random variables. Two decades later, simulations of
different trading mechanisms played an important role in the
literature on the ``microstructure'' of financial markets (Cohen {\em et
al.} \cite{Cohen}). The interest here was mainly in questions of
efficiency and stability of different forms of market organization
and regulation as well as the impact of introducing computer-assisted trading. Like in the approach of Kim and Markowitz a few years later, the sheer complexity of the models, because of the aim to reproduce many features of real-life market,
necessitated a simulation approach. Interestingly, the
microstructure literature later moved on to other questions,
namely, analysis of asymmetric information among traders. Luckily,
Bayesian learning methods allowed to tackle large classes of
asymmetric information models in a rigorous mathematical manner. As
a consequence, the leading textbook of the nineties, ``Market Microstructure Theory'' by O'Hara \cite{OHara}, only reviews theoretical work and lacks any reference to microscopic simulations.

Of course, it was only a matter of time,
until models became so complicated that they could not be solved
analytically anymore and had to be supported by numerical
analysis. In the asymmetric information literature, interesting
recent contributions coming close to microscopic simulations deal with
learning in financial markets (de Fontnouvelle \cite{deFontnouvelle}, Routledge \cite{Routledge,RoutledgeII}). Using different variants of adaptive learning
mechanisms, these authors study how agents learn to use signals
about future market prices and how to make inferences from these
signals. The key interest is in whether or not the learning
dynamics converges to a time-invariant equilibrium that would
obtain under ``rational'' (i.e., correct) expectations.

With its focus on the extraction of information from imperfect signals
by fully rational or learning investors, the dominant branch of models
in financial economics neglected some of the most striking
observations in real financial markets. Namely, there was no role
at all in these models for features like chartist strategies (i.e., strategies looking for patterns in the plots of past prices)
or herd behavior among traders. In some sense, all
traders in traditional microstructure models behave like
fundamentalists  in that they try to infer the correct
``fundamental'' value of an asset from the limited amount of
information they have. However, the existence of both chartists
and fundamentalists in real markets is too obvious to be neglected
and a long tradition of modeling the interaction of these two types
of traders exists in economics literature. In fact, we can find
interesting papers on this subject back in the fifties (Baumol \cite{Baumol}) showing the destabilizing potential of chartist strategies in
a rigorous analytical analysis. The chartist versus fundamentalist topic
was later dropped because of the seeming lack of ``rationality'' of
agents' behavior in these models, that means, the apparent {\em ad-hoc} nature of
the description of individual behavior together with Friedman's \cite{Friedman} argument that irrational traders should be wiped out due to their
incurred losses. While this argument has been often repeated in the economics literature, it is interesting to note that quite a few authors
almost immediately offered
counterexamples to Friedman's assertion \cite{Baumol} \cite{Kemp}. Nevertheless, we can only find
a few contributions to this strand of literature in the seventies and eighties (Zeeman \cite{Zeeman}, Beja
and Goldman \cite{Beja}) and as of the early nineties starting with Day
and Huang \cite{Day} the chartists-fundamentalists interaction regained
its place as an important research topic. The literature of the
nineties has an abundant diversity of interacting agent models
incorporating these features in one or the other way. An early
application to foreign exchange markets is Frankel and Froot \cite{Frankel,Frankel_alt} who
combine a standard monetary model of open economy
macroeconomics with a chartist-fundamentalist approach to expectation
formation (replacing the usual assumption of ``rational''
expectations in earlier models). Their aim is to provide a possible
explanation of the well-known episode of the dollar bubble over the
first half of the eighties. They show that a deviation from the
fundamental value can set into motion a self-reinforcing interplay
between forecasts and actual development: the initial deviation
between price and fundamental value will trigger the switch of some
agents from fundamentalist to chartist behavior. However, the more
the market composition changes in favor of the chartist group, the less
pressure will exist for prices to revert to their fundamental anchor values.
Even laymen far from economic theory or computer simulations may have learned
from the information technology bubble which burst in spring 2000, that not
everything is fully rational in real markets.

An important subsequent variation on
Frankel and Froot's theme is the more elaborate model by DeGrauwe {\em et al.} \cite{DeGrauwe} who show that this type of
dynamics can lead to chaotic behavior of exchange rates. Their
model is one of the first able to explain some stylized facts
other than the mere deviation from the fundamental value. In
particular, they show that their chaotic dynamics is hard to
distinguish from a pure random walk process and that it helps to explain
the forward premium puzzle (the finding, that forward rates are a
poor and biased predictor for subsequent exchange rate movements).

Chaotic dynamics derived from the interaction of agents
with different prediction functions for future price movements are
the topic of a comprehensive research project on ``adaptive belief
systems'' starting with Brock and Hommes \cite{BrockII} and extended in
Brock and Hommes \cite{BrockIII,BrockIV,Brock}, Gaunersdorfer \cite{GaunersdorferI}, Gaunersdorfer and Hommes \cite{GaunersdorferII}, Gaunersdorfer {\em et al.} \cite{GaunersdorferIII}, and Chiarella {\em et al.} \cite{Chiarella} (see also Hommes \cite{Hommes} for a review). While the early papers of this literature are mainly concerned with various bifurcation routes of chaotic attractors
in such systems, the recent papers by Gaunersdorfer and Hommes \cite{GaunersdorferII} and Gaunersdorfer {\em et al.} \cite{GaunersdorferIII} are concerned with
a possible mechanism for volatility clustering emerging from this
theoretical set-up. They show that co-existence of different attractors (e.g., a
fixed point and a cycle or chaotic attractor) in a deterministic
dynamics will lead to repeated switches between these attractors
when small amounts of noise are introduced. Since different attractors
are characterized by different degrees of volatility of prices,
their varying influence on the overall time series generates a
perplexingly realistic picture of switches of the market from
tranquil to volatile phases and {\em vice versa}. Gaunersdorfer and
Hommes \cite{GaunersdorferII} show that estimates of GARCH models can produce quite
similar results as with empirical data.

The adaptive belief dynamics has agents switching between predictors according
to their past performance. A group of alternative learning models
have used modern computer-learning techniques as models of human
adaptation. The best-known variant in the context of financial
markets is surely the Santa Fe  Artificial Stock Market (Arthur {\em et
al.} \cite{Arthur}, LeBaron {\em et al.} \cite{LeBaronIII}, Palmer {\em et al.} \cite{Palmer}),the authors of which included a statistical physicist. In this model, traders are
equipped with a set of classifiers basically consisting of simple
chartist and fundamentalist rules. Particular forecasts of future
returns are connected with certain combinations of classifiers. Classifiers
and forecasts are subjected to genetic operations (selection,
cross-over, mutation). Over time, successful combination of rules
(classifiers) should be maintained, whereas poor ones should be
skipped in favor of better ones. The set-up of this and similar
models is notably different from most other applications of
machine learning techniques: whereas usually classifier systems,
genetic programming, and neural networks are used to recover
regularities in data sets that are independent from their
own learning activity, the artificial financial market application
deals with interacting agents, who naturally influence the
performance of each others' attempt at learning the market's rules.
The main finding of the early work at the Santa Fe Institute was
that the dominance of either chartist or fundamentalist classifiers
depends on the frequency of activation of the genetic operations.
With more frequent activation, chartist behavior was found to be dominating.
LeBaron {\em et al.} \cite{LeBaronIII} showed that the model reproduces some empirical features like leptokurtosis of returns and correlation between
volume and volatility. Other artificial markets include Chen and
Yeh \cite{ChenI,ChenII}, who instead of classifiers systems use genetic programs as evolving models of their agents and also can show
consistency of simulated data with some empirical findings. Cincotti {\em et al.} \cite{Cincotti} construct a more general framework that is designed
to accommodate various learning devices. Related research using
genetic algorithm learning in prominent economic models can be
found in Arifovic \cite{ArifovicII}, Arifovic and Masson \cite{Arifovic}, Dawid \cite{Dawid}, Szpiro \cite{Szpiro}, Lux and Schornstein
\cite{LuxundSchorn} and Georges \cite{Georges}. Le Baron \cite{LeBaronII,LeBaron} has
models closely related to the SFI model, but with learning via
neural networks and the interesting addition of variable memory
length of the agents (cf. the Levy-Levy-Solomon model reviewed
in section \ref{LevyLevySolomon}).

Another strand of economic literature proposed to cope with the
diversity of behavioral variants using a statistical approach cf.
Champernowne and Cowell \cite{Champernowne},
Kirman \cite{Kirman}, Aoki \cite{AokiIII,Aoki_neu}, Ramsey
\cite{Ramsey}, Lux \cite{LuxII}, Foley \cite{Foley} and Kaizoji
\cite{Kaizoji,KaizojiII,KaizojiIII}. Only part of this work is
concerned with financial applications. A wealth of applications of
statistical physics tools to other branches of economics can be found in
Aoki's books \cite{AokiIII,Aoki_neu,AokiYo}.

As concerns finance, perhaps the
first attempt at a microscopic approach with stochastic features guided
by work in statistical physics is Landes and Loistl \cite{Landes}.
Later work includes Youssefmir {\em et al.} \cite{YoussefmirII},
who reconsider the destabilizing potential of trend-following
behavior, and Kirman \cite{KirmanIII} combining the statistical
modeling of herding among speculators with an expectation
formation \`{a} la Frankel and Froot. Similarly, Farmer and Joshi
\cite{Farmer} reconsider the impact of several frequently used
trading strategies in price formation, and Carvalho
\cite{Carvalho} shows that in a particular simplified variant of
their model, emergence of a power-law for extreme returns can be
rigorously demonstrated. Another highly relevant contribution is
Aoki \cite{AokiII} who deals with a stochastic framework for
market participation with infinitely many strategies or trading
rules. Deriving the partition vector (the number of types or
clusters of agents) from a rather general specification of the
entry and exit dynamics, he shows that often the sum of the
fractions of agents in the two largest groups will be close to 1.
This may provide a theoretical rationale for the confinement to
two trader groups in many models of speculative dynamics.

Another example of a statistical approach towards interacting agent
dynamics in finance is the work by Lux and Marchesi, reviewed
below (section \ref{LuxMarchesi}). The latter group of models is,
in fact, not too far from those proposed in the physics
literature. Prominent early examples are the threshold dynamics
(in the form of trigger values for agents' buy or sell decisions)
by Takayasu {\em et al.} \cite{Takayasu} and Bak {\em et al.}
\cite{Bak}. Their analysis is also concerned with scaling behavior
of the resulting price dynamics and reports some interesting
features. A somewhat related model leading to intermittent bursts
of activity is Ponzi and Aizawa \cite{Ponzi}. Later additions to
that literature include the Cont-Bouchaud percolation model
(reviewed in section \ref{ContBouchaud}), and related
lattice-based set-ups by Iori \cite{IoriIII} and Bornholdt
\cite{Bornholdt}. Interestingly, contributions in this vein have
recently also been applied to other financial phenomenon like
contagion of bankruptcies and systemic risk in the inter-bank
lending system (Heymann {\em et al.} \cite{Heymann}, Iori and
Jafarey \cite{IoriII}, Aleksiejuk and Holyst \cite{Aleksiejuk}).

\section{The Dynamics of Monetary Exchange}\label{MonEx}

Before money was invented, exchange of goods would have required barter
between agents with coincident endowments and wants. However, at a more advanced
level of division of labor, one may trade by getting something one
does already possess but which, as judged from past experience,
one will be able to sell later easily to others. 
(The Latin word pecunia for money comes  from pecus = cattle.)
Donangelo and
Sneppen \cite{Donangelo} in this sense started with traders who
initially have a random endowment of products. They then try to
fill the gaps in their inventories by bartering with other
traders, and keep in mind how often some specific product was
asked from them. In the case a trader has something to sell but
already has the product which the partner offers for barter, the
first trader may opt to get a product already in his/her
inventory. This is done with a probability proportional to the
number of times this product was asked from this trader in recent
times, and this product then plays the role of {\em money}: we
cannot eat the money we earn, but we hope to buy food from it
later.

\smallskip

For a suitable range of the number of units per trader and the
number of differentiated products available, traders have enough holes
in their inventories to barter, but after some time also trades
involving money (in the above sense) play an important role; and
sometimes no trade at all is possible in an encounter of two
randomly selected traders. Which product evolves as the most desired
``money'' thus depends on the random dynamics of the market,
without outside interference and without any special property of
that product at the beginning. This result conforms to physics
ideas that ``everything'' can be described by randomness, whether
it is Boltzmann statistics for thermodynamics, the built-up of
social hierarchies \cite{Bonabeau}, or the value of the European
currency. Economists may regard this view as over-simplified.

\smallskip

For one variant of the model, the time-dependence could be
quantified: A stationary state is reached if every trader had
several chances to trade with every possible other trader. The
distribution of times for which one currency stays on top, then
appears to follow a stretched exponential \cite{Stauffer}. Other models for the ``statistical mechanics of money'' are surveyed in \cite{Hayes}.

\smallskip

From the economists' point of view, the informational content of some of these studies is somewhat questionable as there are practically no measurements of the corresponding quantities in real economies. It is nevertheless interesting to note that quite similar models have been brought up by
economists some time ago. Looking up contributions like
the work by Jones \cite{Jones} or the seminal paper by Kiyotaki and Wright \cite{Kiyotaki}, one finds almost the same structure as in the Donangelo and Sneppen
approach. This is not too surprising insofar as - although Donangelo and Sneppen do not quote
the rich literature that emerged from Kiyotaki's and Wright's search model - their work can, in
fact, be traced back to these sources. A careful reading reveals that they draw their
inspiration from an earlier paper in the physics literature, Yasutomi \cite{Yasutomi}, who studied a model
along the lines of Kiyotaki and Wright. It might have been useful to consult the by now
voluminous literature on search-equilibrium models in economics rather than start from scratch
with a similar pursuit. Be that is it may, the style of analysis in the early papers by economists was clearly different from
that of Donangelo and Sneppen. Following the
then prevalent style of reasoning in their subject they
were theoretical investigations into the nature of equilibria in
an economy with a large number of goods rather than truely dynamic model of the emergence of money. The question pursued was
under what conditions one would find a ``monetary'' equilibrium in which
one of the available goods emerges as a medium of exchange and
under what conditions the economy remains stuck in a situation of
barter trade. Like in many other areas in economics, the
demonstration  of existence of multiple equilibria (barter vs.
monetary equilibrium, as well as different monetary equilibria)
pointed to the necessity of investigating out-of-equilibrium
dynamics.

To give the reader a feeling of the typical approach pursued in
economics, we give a short sketch of the basic ingredients of the
seminal Kiyotaki and Wright model that has stimulated a whole
branch of recent economics literature. The set-up by Kiyotaki and
Wright is, in fact, more that of an example than a general model
of a multi-good economy. In particular, it is assumed that there
are three commodities in the economy which are called goods 1, 2,
and 3. There is also an infinite number of individuals who are
specialized in both production and consumption: type $i$
$(i=1,2,3)$ agents derive pleasure (utility) only from consumption
of good $i$, and are able to produce only good $i'$ $\not=$ $i$. A
typical example used in many of the pertinent contributions has
the following structure of consumption and production:
$\begin{array}{| crrr} i & 1 & 2 & 3 \\ \hline
i' & 2 & 3 & 1 \\
\end{array}$ .

This implies that there is no double ``coincidence of wants'' in
the economy. Therefore, intermediate trading of goods by agents
who do not desire them as consumption goods is required for the
satisfaction of the need of these agents. It is furthermore
assumed that in every period there is a random matching process
that assigns every agent to a pair with one other agent within the
economy. Pairs of agents then have the chance to trade with each
other (exchange their goods). In the theoretical papers on this
subject, the focus is on the detection and characterization of
steady state Nash equilibria: sets of trading strategies of each
type of agents together with the steady state distribution of
goods resulting from these strategies, so that each individual
maximizes its expected utility under full information (rational
expectations) about the strategies pursued by other individuals.
There are also storage costs per period for goods that are not
consumed by their owners. The distribution of both the
instantaneous utilities derived from consumption and the storage
costs are crucial for the types of Nash equilibria that exist in
this model. Those goods with lower storage costs are, then, more likely to emerge as ''monneys'' due to their more convenient properties (e.g. shells
rather than pigs). A particular interesting situation is co-existence of
so-called ``fundamental'' and ``speculative'' equilibria. In the
former, only goods with lower storage costs are accepted by the
agents (and, hence, they can be said to concentrate on
fundamentals in their trading decisions) while in the latter case
also some low-storage costs are traded against high-storage
commodities. The motivation for this at first view unattractive
exchange is higher marketability of the high-cost good. Accepting
high-storage costs in the hope of higher chances to exchange these
goods against their preferred one, the agents could be said to act
out of a speculative motivation. This second case is the more
interesting one as it corresponds to the ``emergence of money'':
certain goods are not traded because of their intrinsic values,
but purely because they are accepted by other agents. To solve for
steady state equilibria requires to consider the development of
expected life time utility for each group of agents:

\begin{equation}
E\sum^\infty_{t=0}\beta^t[I^u_i (t) U_i - I^D_{i'} (t) D_i - I^C_{ij} (t) c_{ij}]
\end{equation}
where $U_i$ is instantaneous utility from consumption, $D_i$ instantaneous disutility from
production (i.e., production costs), and $c_{ij}$ the storage costs of good j for type
i. $ \beta<1 $ is the discount factor and $I^u_i$, $I^D_i$ and $I^c_{ij}$ are indicator
functions assuming the value 1 at any period t in which consumption, production or exchange take place and 0 otherwise. Bellman's approach to dynamic programming allows to express this problem in terms of value functions of certain states. For example,

\begin{equation}
V_i(j)= -c_{ij} + \max \beta E[V_i(j')|j]
\end{equation}
could be used to denote  the value for an individual of group $i$ to currently own one unit of good $j$. The value, $ V_i(j) $, of this scenario consists of an instantaneous disutility, $ -c_{ij} $, the negative storage costs incurred by this agent plus the discounted value of the expected change of its situation in the next period, $ E[V_i(j')|j] $. Here  $j'$ could be identical to $j$ (if (s)he does not accept the exchange possibilities offered in the next period), to $i$ (if (s)he is offered her/his preferred good), or some $j' \neq i$ and $j' \neq j$ (if (s)he accepts another good offered to him). Although this formalism greatly facilitates the analysis, rigorous derivation of the type of Nash equilibria sketched above is still a combinatorial nightmare.
Of course, having demonstrated the potential of this kind of model to generate speculative equilibria as steady state solutions, the question emerges whether agents could detect these profitable trading possibilities. A number of authors have taken up the question of whether reasonable dynamics could lead to a self-organization of the Kiyotaki and Wright economy converging to either a fundamental or speculative equilibrium. Contributions to the dynamics of exchange economies made
use of classifier systems \cite{Marimon} or genetic
algorithms in order to describe the evolution of conventions and
self-organization of monetary exchange within an ensemble of
uncoordinated agents.

Besides articles with a computational approach of artificial and
boundedly rational agents one can also find contributions with
real agents in controlled laboratory environments being rewarded
with real money in dependence on their utility gains
\cite{Brown,DuffyI}. To the surprise and disappointment of some
authors, both in experiments with artificial agents \cite{Marimon}
and human subjects one often \cite{Brown,DuffyII} only finds
emergence of fundamental equilibria. Strangely enough a kind of
fundamental steady state even appeared in some set-ups in which
the ``speculative'' scenario is the unique equilibrium. Somewhat
more favorable results concerning the ``emergence of money'' are
obtained in a recent paper by Basci \cite{Basci} who allows for
imitative behavior.

Duffy \cite{DuffyII} tried to combine artificial, agent-based simulations with laboratory experiments with a view to the above mentioned problem. He uses results of preliminary laboratory experiments for his computational approach which leads to an improvement with respect to the speed of learning compared with earlier experimental Kiyotaki-Wright environments \cite{Kiyotaki}.

Furthermore, Aoki \cite{Aoki} uses tools from
statistical mechanics in his re-investigations of the
Kiyotaki-Wright approach. In this perspective, the Donangelo and
Sneppen approach appears to fit well into an established line of
economics research, on the intriguing question:
how could agents develop the idea of ``money''? The early stage of
the study of out-of-equilibrium dynamics in this context warrants
that a great deal of collaborative work could still be done in
this area in the future.

\section{The First Modern Multi-Agent Model: Kim-Markowitz and the Crash of '87 \label{KimMarkowitz}}

After this digression into very fundamental questions of economic
theorizing, we turn to the major playground of multi-agent models
in economics: artificial economic life in the sense of
computer-based stock or foreign exchange markets. Besides some
early Monte Carlo simulations like Stigler \cite{Stigler} or Cohen {\em et al.} \cite{Cohen}, the
first ``modern'' multi-agent model is the one proposed by Kim and Markowitz \cite{Kim}. The major motivation of their microsimulation study was the stock market crash in 1987 when the U.S. stock market decreased by more than twenty percent. Since this dramatic decrease could not be explained by the emergence of significant
new information, ensuing research concentrated on factors other
than information-based trading in determining stock price
volatility (cf. \cite{Schwert}). But although hedging strategies,
and portfolio insurance in particular, have been blamed to have
contributed to the crash by increasing volatility \cite{Brady}, the theoretical work on the link between portfolio
insurance and stock market volatility was rather limited at that time (e.g., \cite{Brennan}). In their simulation analysis, Kim and Markowitz, therefore, tried to explore the relationship between the share of agents pursuing portfolio insurance strategies and the volatility of the
market.

\subsection{The Model}

The simulated market contains two types of
investors, ``rebalancers'' and ``portfolio insurers'', and two assets,
stocks and cash (with interest rate equal to $0$). The wealth {\em w} of each agent at time {\em t} is given as
\begin{equation}
 w_t=q_tp_t+c_t
\end{equation}
where $q_t$ is the number of stocks the agent holds at time $t$, $p_t$ is the price of the
stock at time $t$ and $c_t$ denotes the cash holdings of the agent at time $t$.
{\em Rebalancers} aim at keeping one half of their
wealth in stocks and the other half in cash, i.e.

\begin{equation}
\mbox{target of rebalancers}: q_tp_t=c_t=0.5w_t \mbox{.}
\end{equation}

Thus, the rebalancing
strategy has a stabilizing effect on the market: increasing prices
induce the rebalancers to raise their supply or reduce their
demand; decreasing prices have the opposite effect. {\em Portfolio insurers}, on the other hand, follow a strategy intended to
guarantee a minimal level of wealth (the so-called ``floor'' $f$) at a
specified insurance expiration date. They use the ``Constant
Proportion Portfolio Insurance'' (CPPI) method proposed by Black and Jones \cite{Black}. The method can be described as keeping the value
of the risky asset in a constant proportion to the so-called
``cushion'' $s$, which is the current portfolio value less the floor, i.e.

\begin{equation}
\mbox{target of portfolio insurers}: q_tp_t=ks_t=k(w_t-f)
\end{equation}
where the CPPI multiple $k$ is chosen greater than 1. The floor $f$ is determined as a fraction of the initial wealth and is, therefore, constant over
the duration of the insurance plan. Setting the
multiple above 1 allows the investor to choose her exposure to the
risky asset in excess of  the cushion, and hence to extend her
gains if prices increase. In case of falling prices, the cushion
also decreases and the stock position is reduced accordingly.
Given a more or less continuous revaluation of the portfolio
structure, the floor is therefore (quite) safe. In the presence of falling stock prices, the falling wealth of the investor will eventually approach
his floor from above: since the right-hand side of (5) approaches zero in an extended bear market, the fraction of the riskly asset in the investor's
portfolio will also go the zero. With only riskless cash left in the portfolio, the designated floor, then, constitutes, in fact, the lower bound to the value of
his/her portfolio (if the trading frequency is high enough). In this way, the
Black-Jones formula imitates the effect of put options often
applied in dynamic hedging strategies. Contrary to the rebalancing
strategy, the portfolio insurance strategy implies a procyclical
and therefore potentially destabilizing investment behavior: when
prices fall, portfolio insurers will strive to protect their floor
by reducing their stock position, and conversely, if prices
increase, they will try to raise their stock position in order to
realize additional gains. Note that since $w_t$ on the right-hand side of (5) is mainly governed by capital gains and losses, the multiple $k$ shows
the strength of procyclical reactions of portfolio insurers on price changes.

Stock price and trading volume
evolve endogenously according to demand and supply. However,
trading does not proceed continuously but at discrete points in time. Each investor reviews her portfolio at random intervals. She rates her asset positions using an
individual price forecast computed according to the current
demand and supply situation in the following way:

\begin{enumerate}
\item If only asks (i.e. buy orders) exist, the investor estimates
the price at 101\% of the highest ask price, \item if only open
bids (i.e. sell orders) exist, the investor estimates the price at
99\% of the lowest bid price, \item if both open asks and bids
exist, the investor assumes that the price agreed upon by buyers
and sellers will be placed somewhere between open bid and ask
prices. More precisely it is assumed that her estimate of the new
price is the average between the highest ask and the lowest bid
price of the previous period, and \item if neither asks nor bids
exist, the investor assumes next period's price to equal the
previous trading price.
\end{enumerate}

Summarizing, the above assumptions amount to:

\begin{equation}
p^i_{est,t}=\left\{%
\begin{array}{l}
1.01 \max(p^1_{ask,t}, \cdots ,p^n_{ask,t}), \\
\qquad \mbox{if} \quad \quad p^i_{bid,t}=0 \quad \mbox{for all} \quad i=1,...,n \\
\qquad \mbox{and} \quad p^i_{ask,t}\not=0 \quad \mbox{for at least one} \quad i, \\
\quad \\
0.99 \min(p^1_{bid,t}, \cdots ,p^n_{bid,t}) \quad \mbox{for} \quad p^i_{bid}>0, \\
\qquad \mbox{if} \quad \quad p^i_{ask,t}=0 \quad \mbox{for all} \quad i=1,...,n \\
\qquad \mbox{and} \quad p^i_{bid,t}\neq0 \quad \mbox{for at least one} \quad i, \\
\quad \\
0.5 \left[ \max(p^1_{ask,t}, \cdots ,p^n_{ask,t})+\min(p^1_{bid,t}, \cdots ,p^n_{bid,t}) \right] \\
\qquad \mbox{for} \quad p^i_{bid}>0, \\
\qquad \mbox{if} \quad \quad p^i_{ask,t}\not=0 \quad \mbox{for at least one} \quad i \\
\qquad \mbox{and} \quad p^i_{bid,t}\not=0 \quad \mbox{for at least one} \quad i, \\
\quad \\
p_{t-1}, \quad \mbox{if} \quad p^i_{ask,t}=0 \quad \mbox{and}
\quad p^i_{bid,t}=0 \quad \mbox{for all} \quad i=1,...,n,
\end{array}%
\right.
\end{equation}
where $i$ denotes the agent and $n$ is the number of investors. In
case the estimated ratio between stocks and assets (relevant for
rebalancers) or between stocks and cushion (relevant for portfolio
insurers) is higher than the target ratio (0.5 or $k$ for
rebalancers and portfolio insurers respectively), the investor
will place a sale order with $p^i_{bid,t}=0.99p^i_{est,t}$ (i.e.
$p^i_{ask,t}=0$). Conversely, she will place a buy order if the
evaluated ratio is smaller than the target ratio with
$p^i_{ask,t}=1.01p^i_{est,t}$ (i.e.
$p^i_{bid,t}=0$).\footnote{Strictly speaking, agents allow for
deviations from the target value within a certain range which they tolerate.} If
matching counter-offers exist, incoming buy or sell orders are
executed immediately (at the price of the particular
counter-offer). Otherwise, they are put on a list and may be
filled later during the trading day if suitable offers are made by
other agents. Agents whose orders are open until the end of the
day have the possibility to re-evaluate their portfolio structure
the next day and to place a new order. A trading day is over when
every agent who reviewed her portfolio has had the chance to place
an order and to trade. At the end of each day agents who have lost
their complete wealth (i.e., their cash plus the value of stocks
rated at the closing price) are eliminated and,
thus, excluded from any further trading activities.

\subsection{Results}

Every agent starts the simulation with the same value of her portfolio
(i.e., \$ 100,000), half of it in stocks and half in cash. The
price level at the start of the simulation is \$ 100. The CPPI
multiple $k$ and the insurance level $g$ (i.e., the proportion of
floor to initial assets $f\equiv g\cdot w_0$ with $w_0=p_0q_0+c_0$ denoting the level of initial wealth at time $t=0$) are chosen in a way that portfolio
insurance agents start with their portfolio structure in
equilibrium. The parameters for the insurance plans are set at
(floor / initial assets) = $g =
.75$ (i.e., at expiration date the losses should not exceed 25 \%
of the initial wealth) and $k = 2$. The duration of the insurance
plans is 65 trading days for every plan and each portfolio
insurer. Exogenous market influences are modeled by deposits and
withdrawals of cash occurring at randomly determined points in
time (exponentially distributed with a mean time of 10 trading
days) and in random amounts (uniformly distributed between \$ --8,000
and \$+8,000) for each investor. The time intervals between the
portfolio reviews are also determined by random draws for each
investor (exponentially distributed with a mean time of 5 trading
days).

\begin{figure}
\centering
\includegraphics[width=0.9\textwidth]{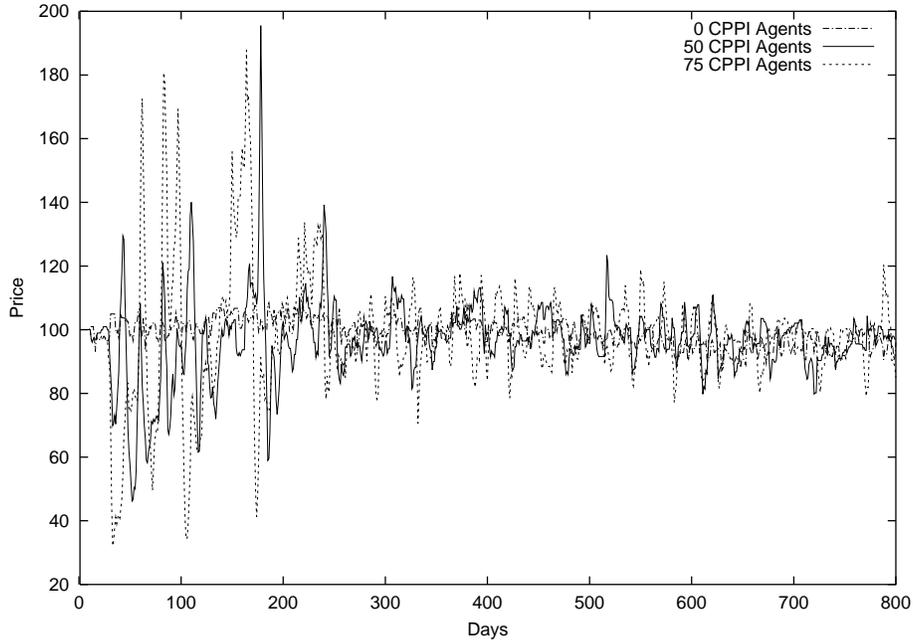}
\caption[]{The daily development of prices (total number of agents: 150)}
\label{fig1}
\end{figure}

\begin{figure}
\centering
\includegraphics[width=0.9\textwidth]{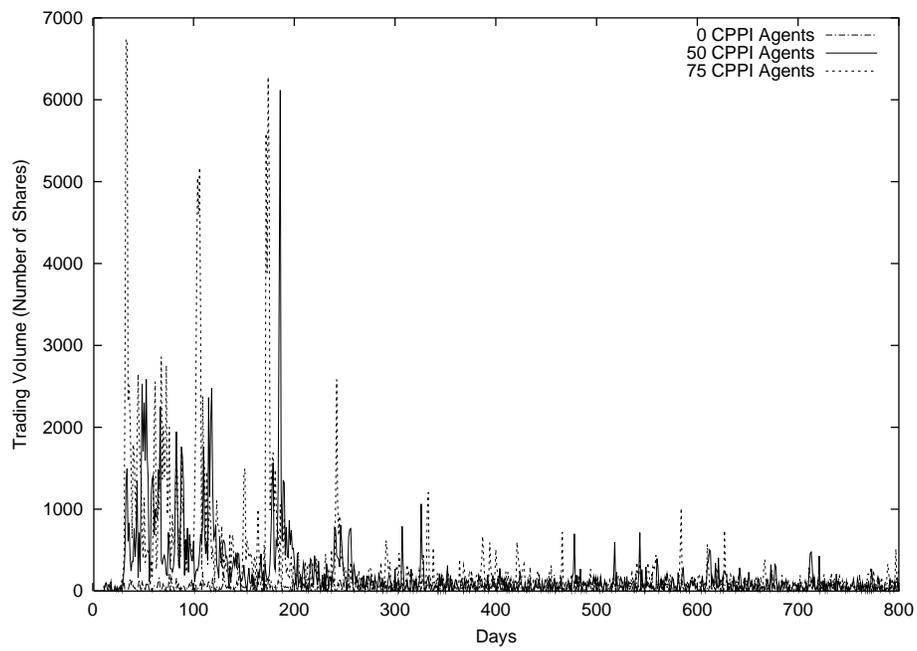}
\caption[]{The daily development of trading volume (total number of agents: 150)}
\label{fig2}
\end{figure}

In the following, we provide the details of simulations in which
we have replicated and extended the results of Kim and Markowitz.
Figure \ref{fig1} and \ref{fig2} show the daily development of
(closing) prices and trading volume for 0, 50 and 75 CPPI agents,
respectively, out of a total of 150 agents for the first 800
trading days. Compared with no CPPI agents, both trading volume
and price fluctuations are generally higher in the cases of 50 and
75 CPPI investors. However, the time series for much more than 50 and 75 CPPI
agents exhibit a cyclical behavior inconsistent with empirical
data \cite{Egenter}.

\begin{figure}
\centering
\includegraphics[width=0.9\textwidth]{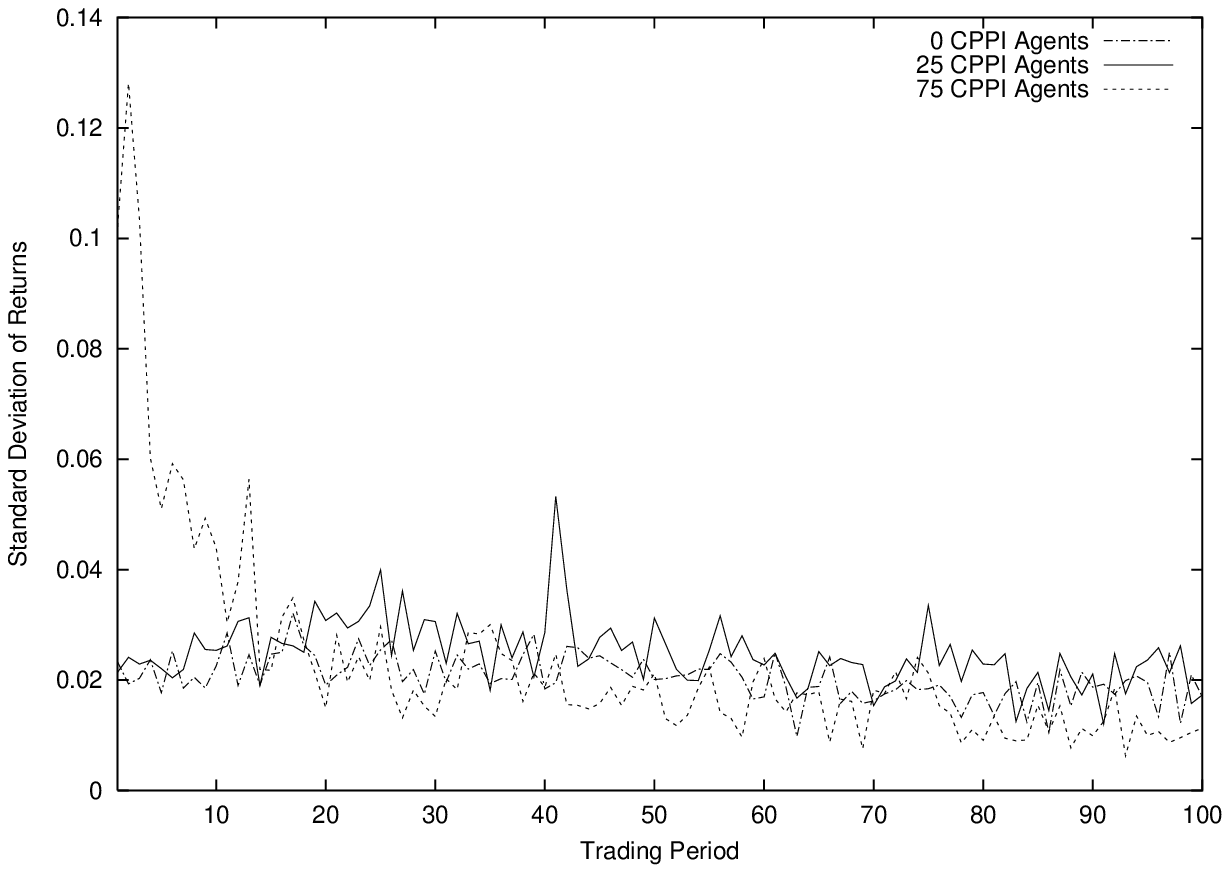}
\caption[]{The standard deviation of daily returns per trading period
(total number of agents: 150)}
\label{fig3}
\end{figure}

We have also studied the standard deviation of daily returns per
trading period each consisting of 65 trading days. As can be seen
in Figure \ref{fig3}, for the first periods the volatility for 75 CPPI
agents is much higher than for 0 or for just 25 CPPI agents. But
after about 15 trading periods the volatility for 75 CPPI agents
declines remarkably. We have observed a similar decline of
volatility in the case of 50 CPPI agents (not displayed in the
figure). The reason, however, for this strong decrease in
volatility in case of a high proportion of CPPI agents is simply
that a significant number of agents become bankrupt in the course of the simulations.

\begin{figure}
\centering
\subfigure{\epsfig{figure=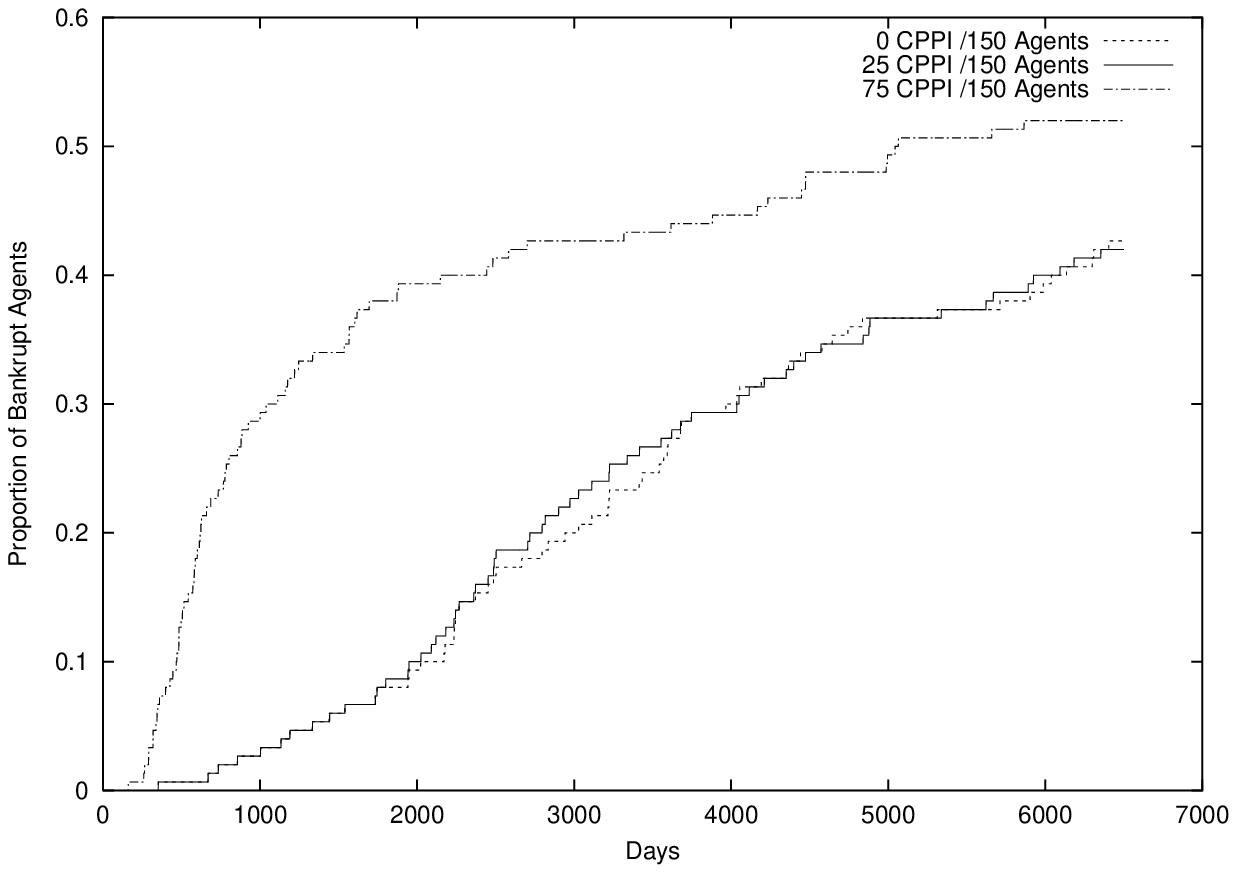,width=0.9\textwidth}}
\quad
\subfigure{\epsfig{figure=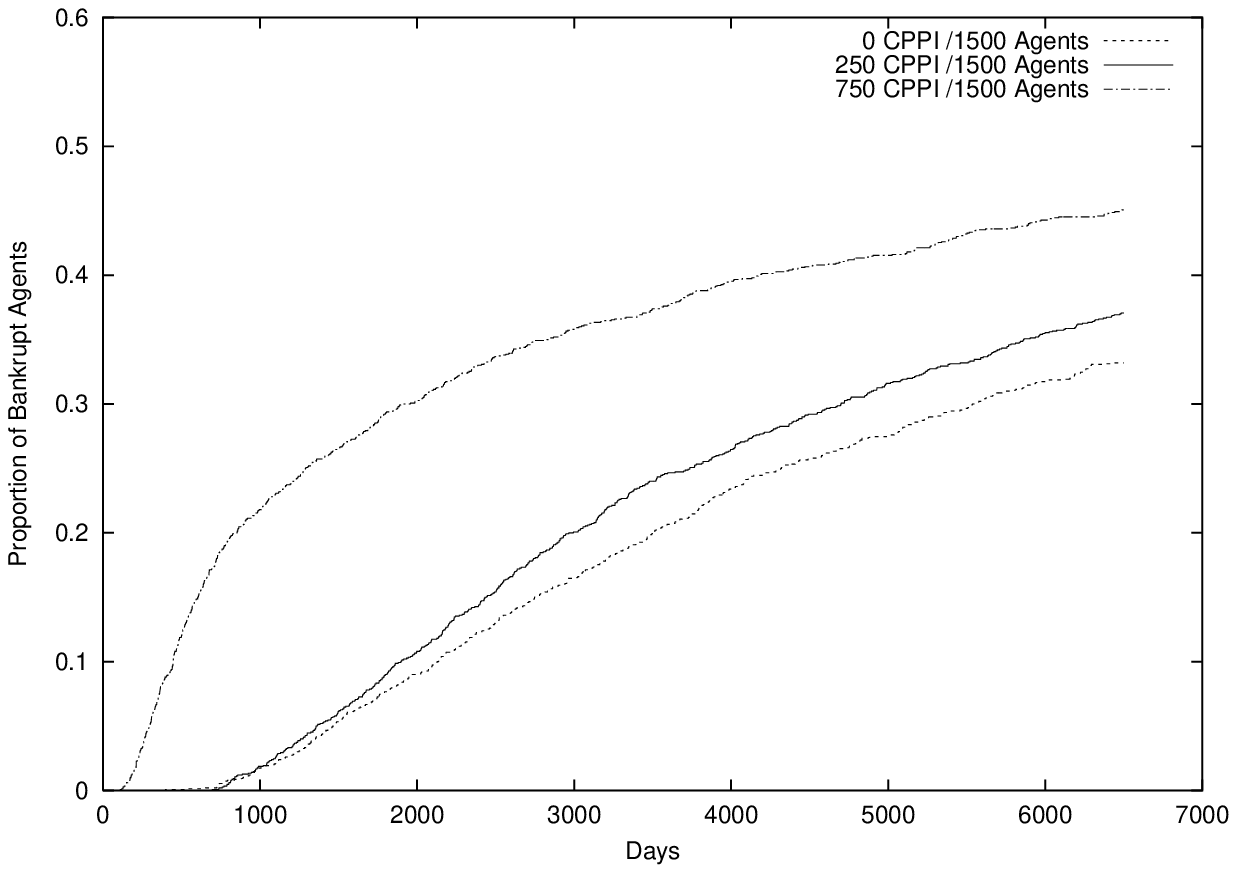,width=0.9\textwidth}}
\caption[]{The proportion of bankrupt investors to the total number of agents (150 and 1500 respectively)}
\label{fig4}
\end{figure}

Figure \ref{fig4} shows a positive relationship between
the proportion of bankrupt agents and the initial share of CPPI
agents. Moreover, there is also a positive relation between the
share of bankrupt CPPI agents in the total number of bankrupt
agents and the initial rate of CPPI agents. Thus, in the case of
25 CPPI agents we had 13 bankrupt CPPI and 50 bankrupt rebalancing
agents after 100 trading periods (i.e., 6500 days), whereas, in the
case of 75 CPPI agents the ratio was 67 CPPI agents to 11
rebalancers (Figure \ref{fig4}, upper panel).

\begin{figure}
\centering
\includegraphics[width=0.9\textwidth]{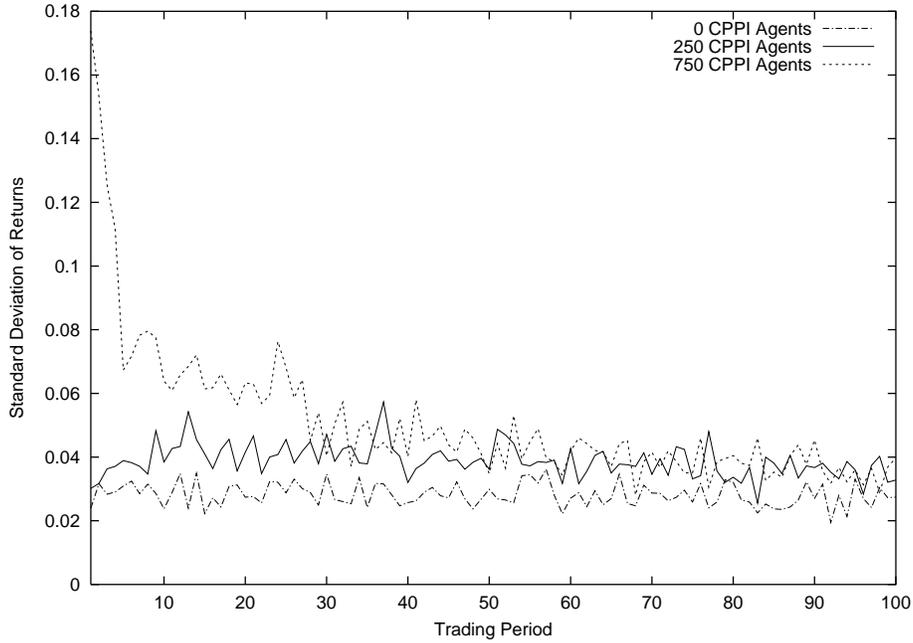}
\caption[]{The standard deviation of daily returns per trading period
(total number of agents: 1500)}
\label{fig5}
\end{figure}

The number of bankrupt investors is generally
lower if we raise the total number of agents to 1500 (Figure
\ref{fig4}, lower panel). Thus, it appears to be a kind of finite-size effect. Nevertheless, in this setting we still observe a reduction of
volatility in the case of a CPPI agents' proportion equal to one
half (i.e., 750 CPPI agents, cf. Figure \ref{fig5}). Compared to the
previous setting, the level of volatility is now significantly higher with
CPPI agents (both 250 and 750 CPPI agents)
than without CPPI agents. From these experiments we conjecture
that the impact of the portfolio insurance strategy on market
volatility generally increases with growing market size.

Nevertheless, given that the model is intended to study the
influence of portfolio insurance on the market, the strong
reduction in the number of active market participants and,
especially, the positive dependence of bankruptcies on the initial
share of CPPI agents, constitutes a serious deficit of the model
design. Presumably however, the quality of the results could be
improved by allowing bankrupt agents to be replaced by new solvent
agents.\footnote{Another way to prevent a large number of bankrupt
agents is to choose an asymmetric distribution for the amounts
withdrawn and deposited on the accounts of the agents. Actually,
by just extending the limit of deposits from \$ 8000 to \$ 9000 (and
keeping the limit for withdrawals at \$ --8000) we discovered a
strong reduction in the number of bankrupt agents.}

\begin{figure}
\centering
\includegraphics[width=0.9\textwidth]{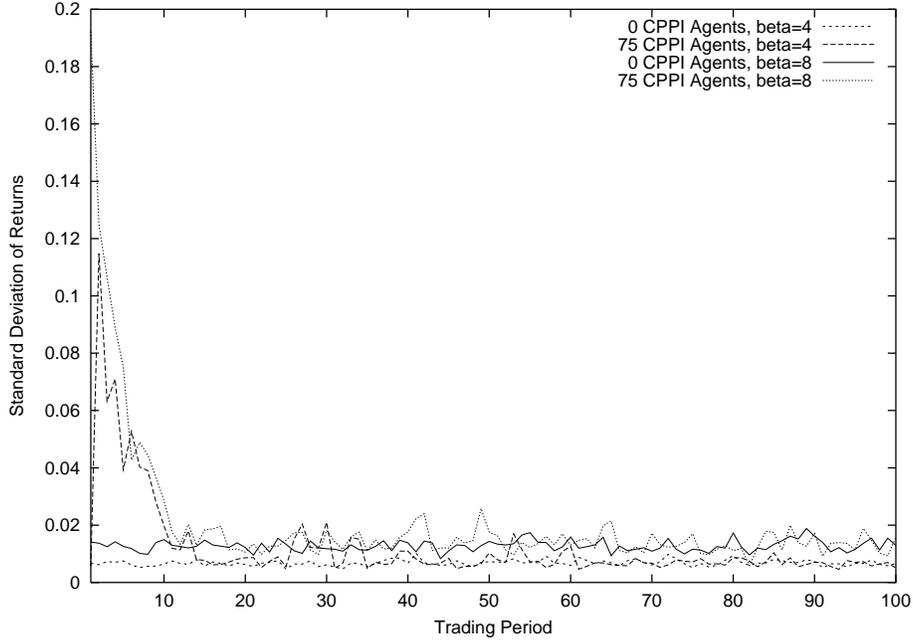}
\caption[]{The standard deviation of daily returns per trading period
for $\beta = 4$ and $\beta = 8$ (total number of agents: 150)}
\label{fig6}
\end{figure}

For a further set of simulations we replaced the individual bid and ask prices by one uniform market price which is set by
a market maker reacting on the difference between supply and
demand.\footnote{A similar modification of the model is described
in Egenter {\em et al.} \cite{Egenter}.} Thus, in case of excess demand (supply) prices rise (fall) proportional to the ratio of excess demand (supply) to the total number of shares with proportionality factor $\beta$:

\begin{equation}
p_t=p_{t-1}(1+\beta \frac{ED_t}{ST_t})
\end{equation}
where $ED$ is the excess demand and $ST$ the total number of stocks in the market. As shown in Figure \ref{fig6}, after about 15
trading periods, the volatility in case of 75 CPPI agents for a
price adjustment speed $\beta = 4$ hardly differs from the case with no
CPPI agents. By increasing the price adjustment speed to $\beta = 8$ the
volatility generally tends to increase (for both 0 and 75 CPPI
agents). As in the
previous setting, in this modified setting the strong decline of
volatility in the case of 75 CPPI agents is again due to the large
number of bankrupt agents.

\begin{figure}
\centering
\includegraphics[width=0.9\textwidth]{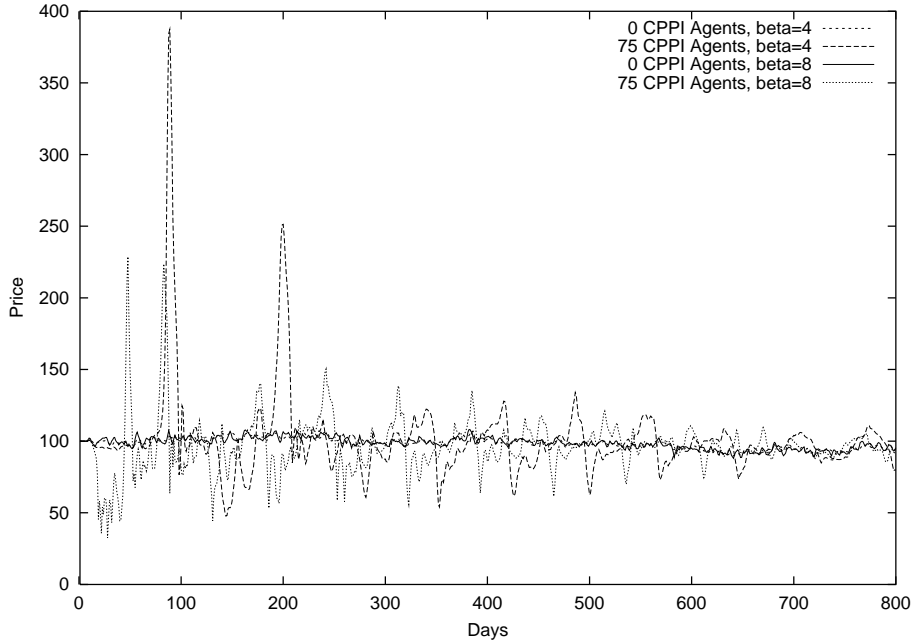}
\caption[]{The daily development of prices for $\beta = 4$ and $\beta = 8$
(total number of agents: 150)}
\label{fig7}
\end{figure}

Also similar to the original setting,
we find almost cyclical price movements for a high proportion of
CPPI agents among our market participants (Figure \ref{fig7}).

\subsection{Conclusions}

Deviating from our parameter setting, the original simulations by
Kim and Markowitz start with the rebalancers' portfolio structure
in disequilibrium, i.e., rebalancers initially have either too many
or too few stocks. Additionally, in their setting, deposits are higher on average than withdrawals. The basic result of this approach
is the demonstration of the destabilizing potential of portfolio
insurance strategies. Kim and Markowitz, therefore, provide a
theoretical foundation for the academic discussions on the sources
of the 1987 crash. Their model, of course, was not designed to
address other puzzles in empirical finance, like the ``stylized
facts'' summarized in the introduction. A comprehensive simulation
study and statistical analysis of model-generated data, in fact,
showed that the time series characteristics exhibit hardly any
similarities with empirical scaling laws \cite{Samanidou}.
Taking into account the pioneering character of this model and the
intention of the authors to provide a partial explanation of the crash of
October '87, our demands on this paper should, however, not be set too high.

\section{An Early 'Econophysics' Approach: Levy-Levy-Solomon}\label{LevyLevySolomon}

Kim and Markowitz obviously tried to simulate a market populated
by traders who pursue strategies found in real-life markets, and,
therefore, gave a quite detailed description of activity at the
microscopic level. In contrast to this highly specific set-up,
more recent models deal with much more stylized and simple
descriptions of traders' behavior. Historically, one of the first of
these approaches is a collaboration of a group at Hebrew
University including both economists and physicists. The first
publication of their approach appeared in an economics journal in
1994 \cite{LevyII} which was followed later by more detailed
reports in physics and computer science journals as well as a
book \cite{LevyIII,LevyIV,LevyV,LevyVI,LevyVII}.

\subsection{The Model Set-Up}

The model contains an ensemble of interacting speculators whose
behavior is derived from a rather traditional utility
maximization scheme. At the beginning of every period each
investor $i$ needs to divide up his entire wealth $W(i)$ into shares
and bonds. Cash, credit, or short sales of stocks are not allowed.
With $X(i)$ denoting the share of stocks in the portfolio of
investor $i$, his wealth can be decomposed as follows:

\begin{equation}
W_{t+1}=\underbrace{ X(i) W_t(i)}_{\mbox{sum of shares}}
+\underbrace{(1-X(i))W_t(i)}_{\mbox{sum of bonds}}
\end{equation}
with superimposed boundaries $0.01<X(i)<0.99.$

\bigskip

Additionally, the model assumes that the number of
investors $n$ as well as the supply of shares $N_A$ are fixed. In addition to an identical utility function $U(W_{t+1})$, investors
at the beginning also possess the same wealth and the same amount of stocks.
Whereas the bond, assumed to be riskless, earns a fixed interest rate $r$,
the stock return $H_t$ is composed of two components (bonds are riskless in economics just like planets are point masses in the first physics lectures).
On the one hand, either capital gains or losses can be the results of price
variations $p_t$. On the other hand,  the shareholder receives a
daily or monthly \footnote{The notion of the underlying time steps differs in the available publications.} dividend payment $D_t$ which grows by a constant rate over time:

\begin{equation}
H_t=\frac{p_t-p_{t-1}+D_t}{p_{t-1}}
\end{equation}

In the base-line model, the preferences of investors are given by
a logarithmic utility function $U(W_{t+1}) = \ln W_{t+1}$. This
function fulfills the usual characteristics of a positively
diminishing marginal utility. The consequence is an absolutely
diminishing risk aversion, so that the amount of money invested in
stocks increases with the wealth of an investor. The so-called
``relative risk aversion'' is constant and the optimal proportion
invested in stocks, therefore, is independent of the wealth (see
Arrow \cite{Arrow}).
The share of stocks, therefore, remains constant.

Investors are assumed to form their expectations of future returns
on the basis of their recent observations. Their memory span
contains the past k total stock returns $H_t$. All investors with
the same memory length $k$ form an investor group $G$. They expect
that the returns in question will reappear in the next period with
a probability of $1/k$. The corresponding expected utility function $EU(X_G (i$)) has to be maximized with respect to the
share of stocks $X_G$:

\begin{equation}
EU = \frac{1}{k} \left[ \sum_{j=t}^{t-k+1}
\ln\left[(1-X_G(i))W_t(i)(1+r)+X_G(i)W_t(i)(1+H_j)\right]\right]
\end{equation}

\begin{equation}
f(X_G(i))= \frac{\partial EU(X_G(i))}{\partial X_G(i)}= \sum_{j=t}^{t-k+1} \frac{1}{X_G(i)+ \frac{1+r}{H_j-r}}= 0
\end{equation}

Like in most models, neither short-selling of assets nor buying
financed by credit is allowed to the agents, so that the space of
admissible solutions is restricted to a share of stocks in the
interval $[0,1]$. Levy, Levy, Solomon, furthermore, impose minimum
and maximum fractions of shares equal to 0.01 and 0.99 in cases
where the optimal solution of the optimization problem would imply
a lower (higher) number. We, hence obtain either inner or outer
solutions for $X_G(i)$ which are depicted in Table 1. (Because of the
maximum condition the first derivative of $f$ is zero.)

\begin{table}
\begin{center}
\textbf{ \ Table
1.} Inner and outer solutions\\
\begin{tabular}{ccc}
\hline $ f(0) $ & \hspace{1cm}$ f(1) $ & \hspace{0.5cm} $X_G(i)$
\\\hline $<0$ & \hspace{1cm}$- $ & $0.01 $ \\\hline $>0$ & \hspace{1cm}$<0$ & \hspace{0.5cm}
$0.01<X(i)<0.99$
\\\hline $>0$ & \hspace{1cm}$>0$ & $0.99$ \\\hline
\end{tabular}\\
\end{center}
\end{table}

When the optimum share of stocks is calculated for an investor
group $X_G(i)$, a normally distributed random number
$\varepsilon_i$ is added to the result in order to derive each individual investor's demand or supply. This stochastic component
may be interpreted as capturing the influence of idiosyncratic
factors or of individual deviations from utility maximization from
the economists' point of view. However, in the original papers it
is motivated from a physics perspective as the influence of the
``temperature'' of the market.  From aggregation of the stochastic
demand functions of traders, the new stock price (and therefore,
the total return $H_t$), can be calculated as an equilibrium
price. One now eliminates the ``oldest'' total return from the
investors' memory span and adds the ``new'' entry when the simulation
process is finished for period $t$.

\subsection{Previous Results}

\begin{figure}
\centering
\includegraphics[width=0.9\textwidth]{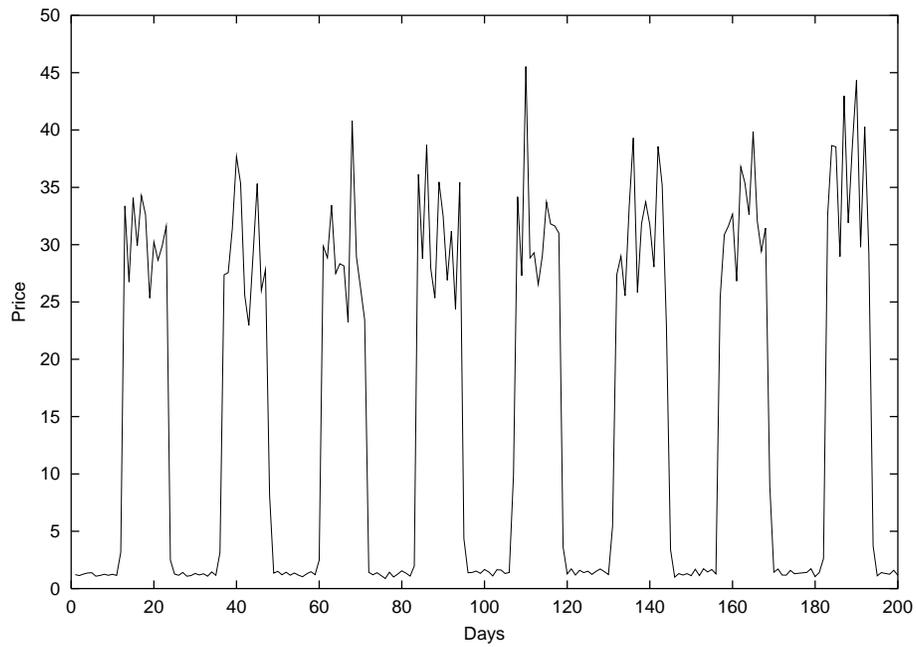}
\caption[]{With only one type of traders and a logarithmic utility
function, the typical outcome of the Levy, Levy, Solomon model is
a cyclic development of stock prices with periodic booms and
crashes. Our own simulations produced all the visible patterns
emphasized in \cite{LevyII,LevyIII,LevyV}.} \label{fig8}
\end{figure}

Models with only one investor group show periodic stock price
developments (Figure \ref{fig8}) whose cycle length depends on the memory span
$k$. This price development can be explained as follows: Let us
assume that, at the beginning of the simulation, a random draw of
the $k$ previous total stock returns $H_t$ occurs that encourages
investors to increase the proportion of shares held in their
portfolio. The larger total demand, then, causes an increase in
price and therefore a new positive total return results. According
to the updating of data the oldest total return will be dropped.
This positive return causes the investor group to raise their
stock shares successively up to a maximum of 99\%. At this high
price level the price remains almost constant for a little longer
than $k$ periods until the extremely positive return of the boom
period drops from the investors' memory span.

As explained above, the total return is composed of the capital
gains or losses and of the dividend. Since the dividend yield, $D/p$, is
relatively small because of the considerably high stock price, a
relatively small (negative) total stock return (caused by the
noise term $\varepsilon_i$) suffices to make the riskless bond
appear more attractive. The desired share of stocks and with it
the stock price, then, break down. If such a crash happens with an
ensuing extremely negative total return the desired share of
stocks drops to a minimum of 1\%. Again, it takes another $k$
periods for the investors to forget about this extremely negative
entry. Because of the then available high real dividend rate,
investing in shares becomes more attractive compared to bonds. The
total demand and the stock price start rising again and a new price
cycle begins. If two groups with different memory spans are
considered, strict periodicity still remains a possible
outcome. However, depending on the choice of the memory spans, other
dynamic patterns can appear. Looking at the distribution of total
wealth, a dominating influence on the share price development by one group then
does not necessarily mean that it also gains a dominant share of total wealth.

\begin{figure}
\centering
\includegraphics[width=0.9\textwidth]{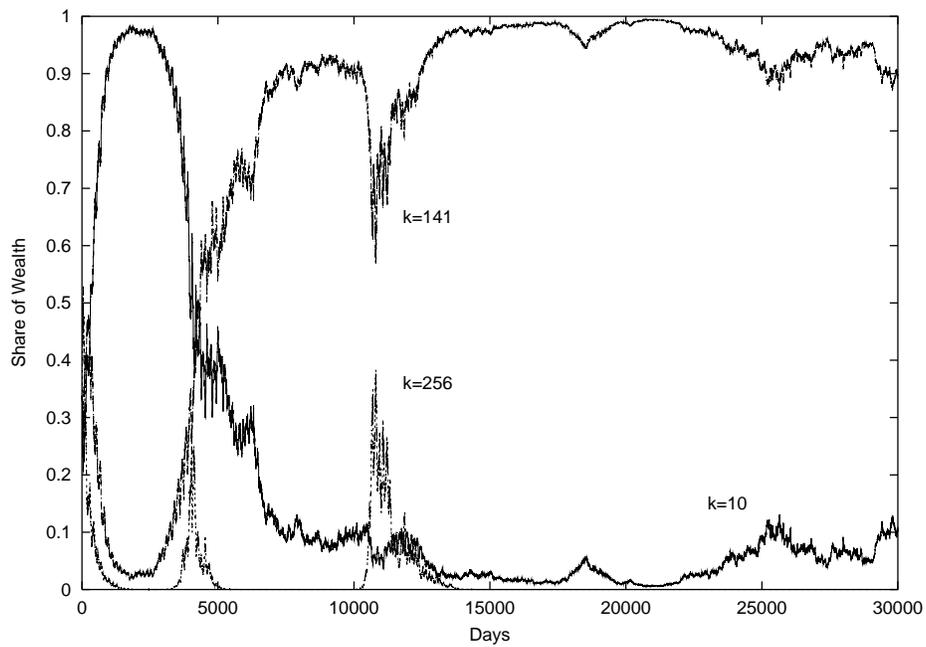}
\caption[]{Development of the distribution of wealth with three groups
characterized by $k = 10$, 141, and 256, respectively. Depending on the
initial conditions, either the group with  $k = 256$ or the group
with $k = 141$ as in the present case may happen to dominate the market}
\label{fig10}
\end{figure}

The model outcome becomes more irregular with three (and more)
investor groups. For example, for the combination $k = 10$, 141, 256
and $n = 100$, Levy and Solomon claimed to have found chaotic motion in stock
prices. However, Hellthaler \cite{Hellthaler} has shown that if the number of investors is increased from $n = 100$ to, for
example, $n = 1000$, these chaotic stock price developments are
replaced by periodic motion again. This effect persists if more than one type of stocks is traded \cite{Kohl}. Furthermore, Zschischang and
Lux \cite{Zschischang} found that the results concerning the wealth
distribution for $k = 10$, 141 and 256 are not stable. While Levy,
Levy and Solomon argued that the group with $k = 256$ usually turned
out to be the dominant one, it is also possible that the investor
group $k = 141$ will achieve dominance (Figure \ref{fig10}). This shows an
interesting extreme type of dependence of the model outcome on
initial conditions brought about by seemingly minor differences
within the first few iterations: depending solely on the random
numbers drawn as the ``history'' of  the agents at $t = 0$, we get
totally different long-run results for the dynamics.

Of course, one would like to have microscopic models to provide an
explanation of the power-law behavior of both large returns and
the time-dependence in various powers of absolute returns.
However, when investigating the statistical properties of Levy,
Levy and Solomon's model, the outcome is as disappointing as with
the Kim and Markowitz framework: none of the empirical scaling
laws can be recovered in any of our simulations (see Zschischang \cite{Zschischang_dipl} who investigates about 300 scenarios with
different utility functions, memory spans and varying number of
groups). Scaling laws have, however, been reported in related models of the same group of authors, e.g. \cite{Levy}.
While the underlying philosophy of both approaches is somewhat similar, their microscopic structure is quite different.

\begin{figure}
\centering
\includegraphics[width=0.9\textwidth]{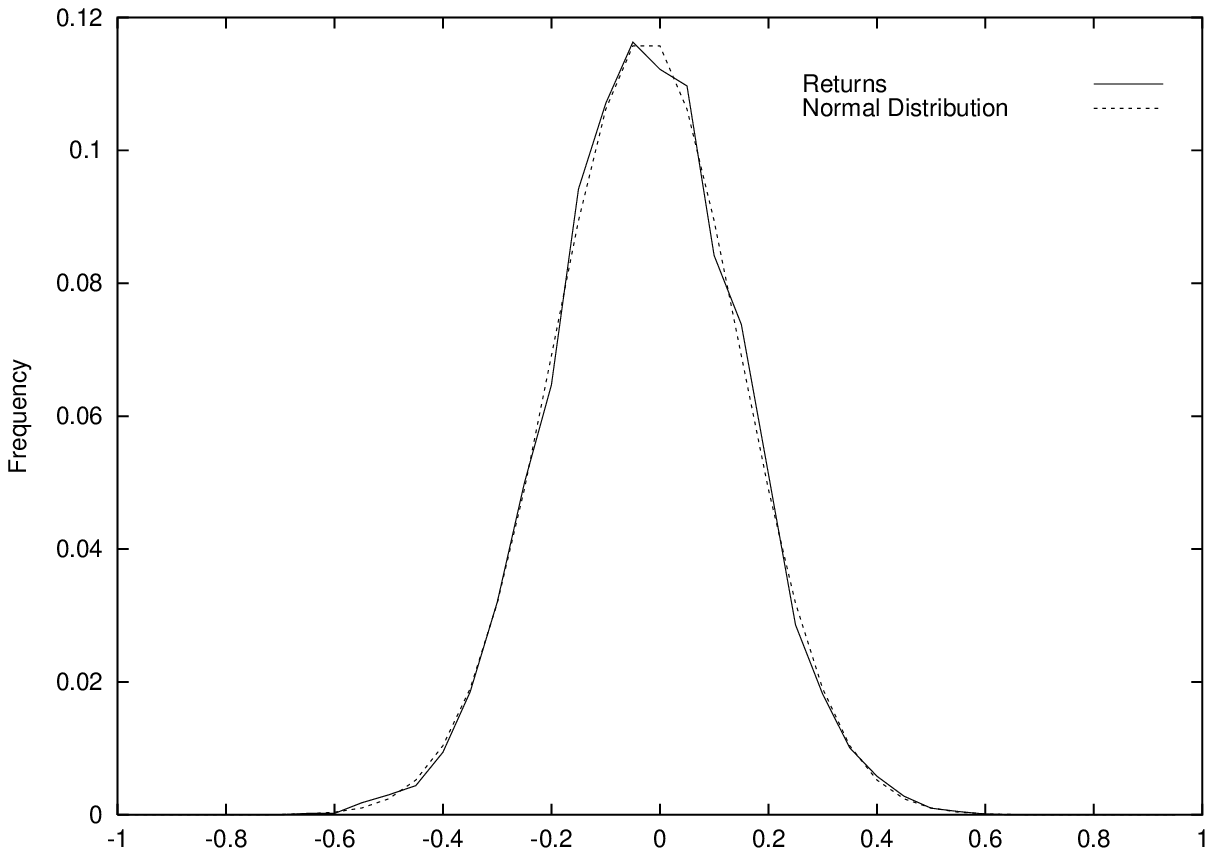}
\caption[]{Distribution of returns form a simulation with 6 groups
characterized by memory spans $k = 10$, 36, 141, 193, 256, and 420. This
is an example with a stock price development described as ``chaotic'' in
{\cite{LevyV}}. However, it seems that the result is rather similar to pure randomness. The histogram is drawn for 20,000 observations after an
initial transient of 100,000 time steps. The close similarity to the
Normal distribution is confirmed by statistical measures: Kurtosis is
0.043 and skewness is -0.003. This yields a Jarque-Bera statistic of
1.55 which does not allow to reject the Normal distribution
(significance is 0.46\%)}
\label{fig11}
\end{figure}

\begin{figure}
\centering
\includegraphics[width=0.9\textwidth]{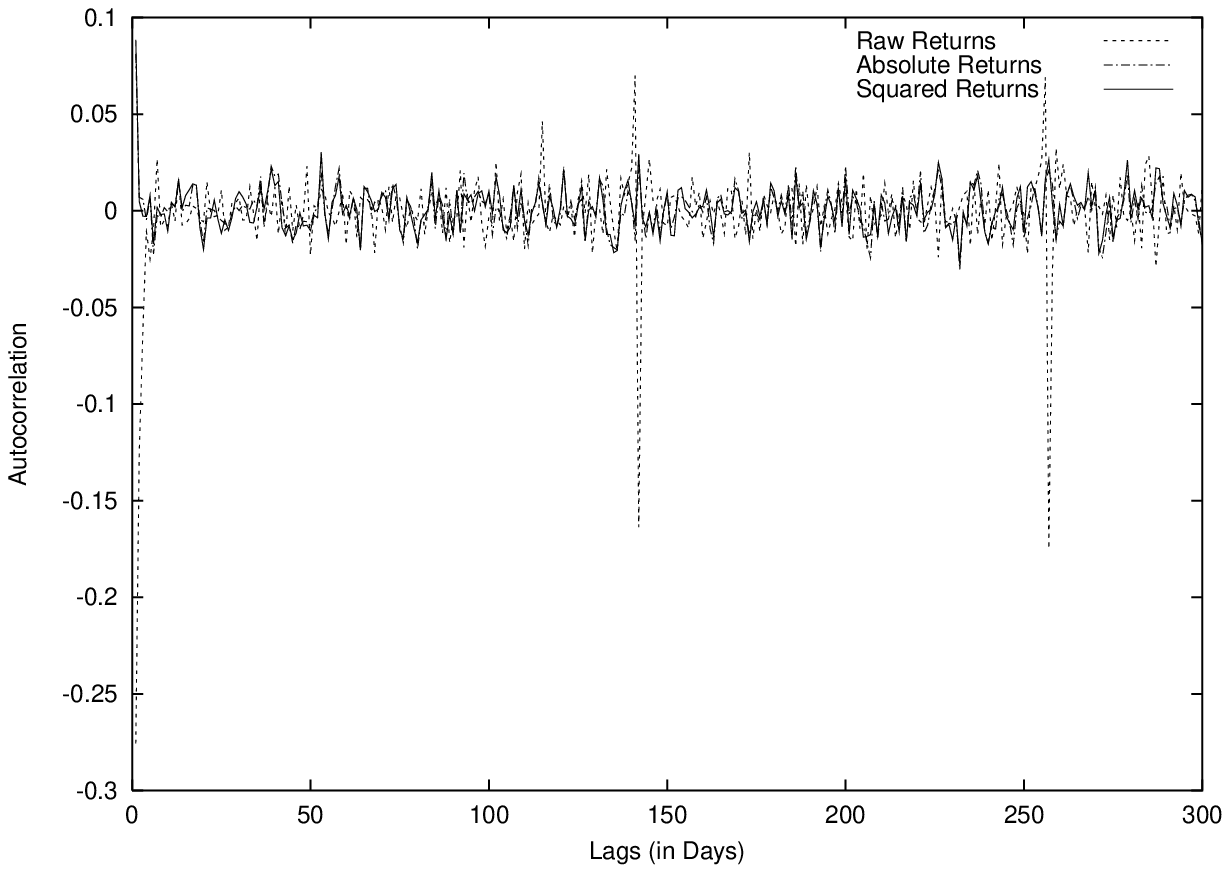}
\caption[]{Autocorrelations of raw, absolute and squared returns from the simulation of a ``chaotic'' case of the Levy-Levy-Solomon model. As can be seen, dependence in absolute and squared returns (as typical proxies for market volatility) is as weak as with raw returns themselves. The underlying scenario is the same as with Figure \ref{fig11}}
\label{fig12}
\end{figure}

As exemplified in Figure \ref{fig11}, models which are
claimed to have a chaotic price development often have stock
returns that appear to follow a Normal distribution (Figure \ref{fig11}) and do
not account for ``clustered volatilities'' (Figure \ref{fig12}). These results are supported by standard tests of the Normal
distribution and for the absence of correlations of stock returns.
Zschischang and Lux argue that in these cases the Levy, Levy,
Solomon model, instead of giving rise to low-dimensional chaotic
dynamics and strange attractors, can effectively be viewed as a
random number generator \cite{Zschischang}.

The original papers are not entirely clear about the lengths of the time increments of the model: they are sometimes denoted by ``days'' and sometimes by ``months''. Since at low frequencies, returns in real markets seem to approach a Gaussian distribution, in such an interpretation, the Normality of returns generated from the model might even appear to be a realistic feature. However, the mechanism for the emergence of a Gaussian shape is still different from its origin in monthly returns in reality. The latter seems to be the consequence of the aggregation of high-frequency returns whose distribution is within the domain of attraction of the Normal distribution (because of its power-law exponent above 2). In the LLS model, on the other hand, the Gaussian shape seems to originate from the aggregation of random demand functions {\em within} the same period.

\section{Financial Markets and the Distribution of Wealth: Solomon-Levy-Huang}\label{SolLevyHuang}

Again, the unrealistic time series characteristics of both the Kim
and Marko\-witz and the Levy, Levy, Solomon approach should not be
taken too seriously: both models are among the first attempts at
microscopic simulations of financial markets and their aims were
more to provide mechanisms for bubbles and crashes than to look at
statistical features of the so generated time series. At least in
the case of Levy, Levy, Solomon, the authors initially were not
aware at all of the scaling laws characterizing financial markets
(personal communication by Sorin Solomon). However, later on their
model served as inspiration and starting point for the analysis of
statistical properties of simulated data. As an interesting
example, the inherent wealth dynamics in Levy, Levy, Solomon
inspired a more thorough analysis of the development of traders'
wealth in some kind of generalized Lotka-Volterra systems.

This extension is based on a 1996 model \cite{Levy,Solomon} which was
re-investigated more recently. The pertinent results have been
presented in a series of recent papers by Huang, Solomon and
others. We thus call it the SLH model. Its mechanics can be
described as a random multiplicative process for the wealth of
each trader, with different traders coupled through their average
wealth somewhat similar to predator-prey models.

Assume that all traders start with the same wealth but later
each of them speculates differently on the market and gains or
loses amounts proportional to his current wealth:

\begin{equation}
w_i(t+1)=\lambda w_i(t) \quad (i = 1,2, \dots, N)
\end{equation}
where $\lambda$ is a number fluctuating in a small interval $D$
centered about unity. This random multiplicative process has been
discussed before. The new ingredient in SLH is the ``welfare state'': Nobody
is allowed to fall below some poverty level $w_i=qW$ where
$W=W(t)=\sum_iw_i(t)/N$ is the average wealth per trader at that
time. Thus this model is very simple, but nevertheless possesses
many realistic properties. Physicists can identify it with a
random walk on a logarithmic wealth scale with one repelling
boundary.\footnote{This passage, having been contributed by D.S.,
obviously reflects the tendency of physicists to know everything
better. In fact, the remaining authors (although they are only
economists) see no reason why they should be unable to recognize a
random walk with a reflecting boundary.} Instead of this cut-off, the authors also
investigate the rule

\begin{equation}
w_i(t+1)=\lambda w_i(t)+ a W(t),
\end{equation}
which represents a rich society engaging in even redistribution of a certain fraction $a$ of overall wealth.

In the infinite $N$ limit the same relative
wealth distribution

\begin{equation}
P\left(\frac{w_i}{W}\right) \propto \left(\frac{w_i}{W}\right)^{-2-2a/D}\exp\left[\frac{-aW}{Dw_i}\right]
\end{equation}
is obtained \cite{SLH7} for a more general and realistic model:

\begin{equation}
w_i(t+1)=\lambda w(i)+a W(t)-c(W,t)w_i(t)
\end{equation}
where the arbitrary function $c(W,t)$ parameterizes the general
state of the economy: time periods during which $-c(W,t)$ is large
and positive correspond to boom periods while periods during which
it is negative correspond to recessions. Complementarily, if one
thinks of $w_i$ as the real wealth (as opposed to the nominal
number of dollars which could increase solely because of
inflation) of each individual, an increase of the total amount of
dollars in the system $W(t)$ means that an agent with individual
wealth $w_i$ will suffer from a real loss due to inflation in an
amount proportional to the increase in average wealth and
proportional to one's own wealth: $-c(W,t)w_i(t)$.

The following results are obtained: For infinite markets, a power law $\propto 1/w^\alpha$ was
obtained for the probability of traders with wealth larger than
some arbitrary wealth $w$. The exponent for this
power law is given by the cut-off: $\alpha=1/(1-q)$. $q$ was defined above
as the ratio of the lowest allowed wealth to the average wealth.  Thus
if $q\cong\frac{1}{3}$ we have $\alpha\cong 1.5$ in
agreement with well-known empirical findings. It would be interesting to see if in
real economies this exponent and the analogous one for the price
fluctuations depend on the lower cut-off for wealth: The more
egalitarian or socialist a country is, the higher will be
$q$, and the higher will be the exponent $\alpha$,
making extreme wealth inequalities and market fluctuations less
probable.

The amount trader $i$ invests on the stock market is proportional
to the wealth $w_i$ : George Soros produces more price changes
than the present authors together (we expect this to change in the
near future). Thus the fluctuations of the market price were at
first thought to have the same tail exponent
$\alpha$ as the wealth distribution. However, this is not
true because the different traders are not statistically
independent: the cut-off $w_i\ge qW$ introduces a coupling via the
average wealth $W$.

Moreover, real markets are finite, and according to a 1999 review
of  microscopic market models \cite{StaufferII}, the majority of these
models get unrealistic properties like periodic oscillations, if
the market size goes to infinity. In short, a few hundred
professional speculators and not the millions of non-speculative
families dominate most of the market movements. The thermodynamic
limit, so traditional in statistical physics where a glass of beer
contains $10^{25}$ molecules and where $5\times10^{13}$ sites
were already simulated \cite{Tiggemann}, may, therefore, be very unrealistic for
markets or social science \cite{Toral}.

Indeed, simulations of the SLH model for $10^2 \ldots 10^4$
traders gave effective exponents $\alpha \simeq 3$, i.e.  close to the desired
one for the price flcutuations (not the wealth distribution).
These exponents are valid only in some intermediate
range: For small wealth the cut-off is important, and nobody can
own more wealth than is available in the whole market. We refer to
the SLH papers for more details on this approach \cite{Huang,SLH2,SLH3,SLH4,SLH5,SLH6,SLH7,SLH8}.

A somewhat related recent strand of literature has analysed simple monetary exchange models.
The main question pursued in this area is emergence of inequality within a pool of agents due to the randomness of their
fortunes in economic interactions. This line of research is represented, among others by
\cite{Bouchaud}, \cite{Chakraborti2}, \cite{Dragulescu,Bennati}. The structure of all these models is very simple: agents are randomly matched in pairs and try to catch
some of the other agent's wealth in this encounter. A random toss decides which of both opponents is the winner of this match.
The successful agent, then, leaves the battle field with her/his wealth having increased by a fraction of the other agent's previous belongings. The
above papers show that this simple random exchange model (with only minor differences in the stochastic formalisation in the above papers)
leads to an endogeneous emergence of inequality  within an initially homogenous population. It is, however, worthwile
to point out that exactly the same process had already been proposed in \cite{Angle86} by sociologist John Angle and has been extended in
various ways in the pertinent literature over the years (\cite{Angle92,Angle96,Angle06}). Needless to say that physicists would have gained
by first consulting the literature on the subject before starting to duplicate well-established lines of
research. It might also be remarked that in the recent economics literature, a number of more realistic models of wealth formation and
agend-based models exist (e.g. \cite{Silver}).
A more extensive discussion of this class of model can be found in \cite{lux05}.

\section{Percolation Theory Applied to Finance: Cont-Bouchaud}\label{ContBouchaud}

Together with the random walk model of Bachelier \cite{Bachelier}
hundred years ago, and the random multiplicative traders of SLH,
the Cont-Bouchaud model is one of the simplest models, having only
a few free parameters (compared, e.g., to the ``terribly complicated"
Lux-Marchesi model reviewed below). Economists like biologists may
bemoan this tendency of physicists, but the senile  third author from
the physics community likes it. Also, it is based on decades of
percolation research in physics, chemistry and mathematics, just
as Iori's random-field Ising model uses many years of physics
experience in that area \cite{Iori}. Obviously, with this type of
models, ``econophysicists'' have introduced new tools of analysis
to financial modeling. As recent research in economics has focused
on communication and information among traders (e.g.,
\cite{Kirman,Banerjee}), the random-field and percolation models
might be welcome means for the investigation of information
transmission or formation of opinions among groups of agents.

In percolation theory, invented by the later chemistry Nobel
laureate Paul Flory in 1941 to explain polymer gelation (cooking
of your breakfast egg), and later applied by Broadbent and
Hammersley to coal-dust filters, and by Stuart Kauffman to the
origin of life, each site of a large lattice is either occupied
(with probability $p$), or empty (with probability $1-p$).

Clusters are groups of occupied neighbors. A cluster is infinite
if its mass $s$ (the number of occupied sites belonging to that
cluster) increases with a positive power of the lattice size (and
not only logarithmically). When the density $p$ increases from
zero to one, at some sharp percolation threshold $p_c$ for the
first time (at least) one infinite cluster appears; for all
$p>p_c$ we have exactly one infinite cluster for large and not too
anisotropic lattices, filling a positive fraction $p_\infty$ of
the whole lattice. For all $p<p_c$ we have no infinite cluster. At
$p=p_c$ we find incipient infinite clusters which are fractal. If
$p$ approaches $p_c$ from above, the fraction $p_\infty$ vanishes
as $(p - p_c)^\beta$ with some critical exponent $\beta$ varying
between zero and unity for dimensionality going from one to
infinity. The Hoshen-Kopelman and the Leath algorithm to produce and
count percolation clusters are well documented with complete programs
available from DS.

The average number $n_s$ of clusters containing $s$ sites each
follows a scaling law for large $s$ close to the threshold $p_c$

\begin{equation}
n_s=s^{-\tau} f[(p - p_c)s^\sigma]
\end{equation}
where the exponent $\tau$ varies from 2 to 2.5 if $d$ goes from
one to infinity, and the exponent $\sigma$ stays close to 1/2 for
$2\le d \le \infty$. The previous exponent $\beta$ equals $(\tau -
2)/\sigma$. The details of the lattice structure do not matter for
the exponents, only for the numerical value of $p_c (\cong
0.5927464$ for nearest neighbors on the square lattice). On a
Bethe lattice (Cayley tree),
$\tau=\frac{5}{2},\, \sigma=\frac{1}{2},\, \beta=1$, and the above
scaling function $f$ for the cluster numbers is a Gaussian; these
exponents are also found for $6<d\le\infty$ and for the case that
each site is a neighbor to all other sites (i.e., a ``random
graph'') \cite{Odagaki}.

The above model is called site percolation; one can also keep all
sites occupied and instead break the links between neighboring
sites with a probability $(1-p)$. This case is known as bond
percolation and has the same exponents as site percolation.
Computer programs to count percolation clusters were published by
many, including the senile co-author
\cite{StaufferIII,Stauffer_alt}. All this knowledge was available
already before percolation was applied to market fluctuations.

In the Cont-Bouchaud market model, originally restricted to the
mathematically solvable random graph limit and later simulated, as
reviewed in \cite{StaufferIV}, on lattices with $2\le d \le 7$, each
occupied site is a trader. A cluster is a group of traders making joint decisions; thus the model
simulates the herding tendency of traders. At each time step, each
cluster either trades (with probability $2a$) or sleeps (with
probability $1-2a$), and when it trades it either buys or
sells an amount proportional to the size $s$ of the cluster $i$;
thus $a$ usually is the probability for the member of a cluster to be a buyer. The market price is driven by the
difference between the total supply and demand; the logarithm of
the price changes proportionally to this difference (or later
\cite{StaufferV,Zhang} to the square-root of the difference). The
concentration $p$ is either fixed or varies over the whole
interval between zero and unity, or between zero and $p_c$. The results deteriorate \cite{Chakraborti,Changetal}
if the price change is no longer proportional
to the difference between demand $D$ and supply $S$, but to the relative
difference $(D-S)/(D+S)$ or to a hyperbolic tangent tanh[const$\cdot(D-S$)]. However, the latter has been found to be a more realistic description of the price impact of demand variations \cite{Plerou}.

\begin{figure}
\centering
\includegraphics[angle=-90,width=0.9\textwidth]{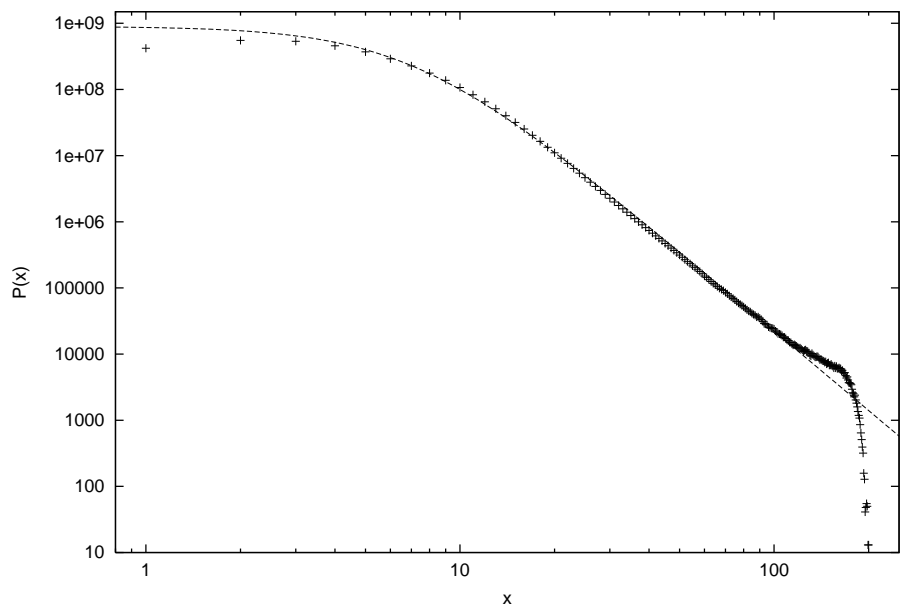}
\caption[]{Distributions of price changes from the Cont-Bouchaud percolation model. The figure also compares $P(x)$ with const/$(3 + 0.06*x^2)^2$. The underlying data are averages from many $301 \times 301$ square lattices}
\label{lux1}
\end{figure}

Some results are: If the activity $a$ increases from small values
to near its maximum value 1/2, the histogram of the price
fluctuations changes from an asymptotic power law to something
close to a Gaussian, similar to the crossover in real markets when
the observation time goes to infinity. For small activities, the
cumulative probability to have a change by at least $x\%$,
varies in its tail as $1/x^\mu$, with \cite{StaufferV,StaufferIV}
$\mu=2(\tau+\sigma-1)$ if we use Zhang's square-root law and
average over all $p$ between 0 and $p_c$. This exponent $\mu$ varies from 2.9 and 3.3 to 4 if $d$ increases from two and three to
infinity. Thus in the realistic dimensions of a city or bank
building, $d=2$ or 3, we get the desired $\mu\cong 3$. Figure \ref{lux1}
shows simulations giving this power law, except for flattening at
small $x$ and a cut-off due to finite market sizes at large $x$. The
curve through the data is the Student-t distribution following
from Tsallis statistics \cite{deSouza}.

Volatility clustering, positive correlations between trading
volume and price fluctuations,  as well as the observed asymmetry
between sharp peaks and flat valleys is seen if the activity
increases (decreases) in a time of increasing (decreasing) market
prices. Nearly log-periodic oscillations are seen if a non-linear
restoring force (buy if the price was low) is combined with some
hysteria (buy if the price was rising). For more details we refer
to the original papers following \cite{Cont} reviewed in \cite{StaufferIV}.

\begin{figure}
\centering
\includegraphics[angle=-90,width=0.9\textwidth]{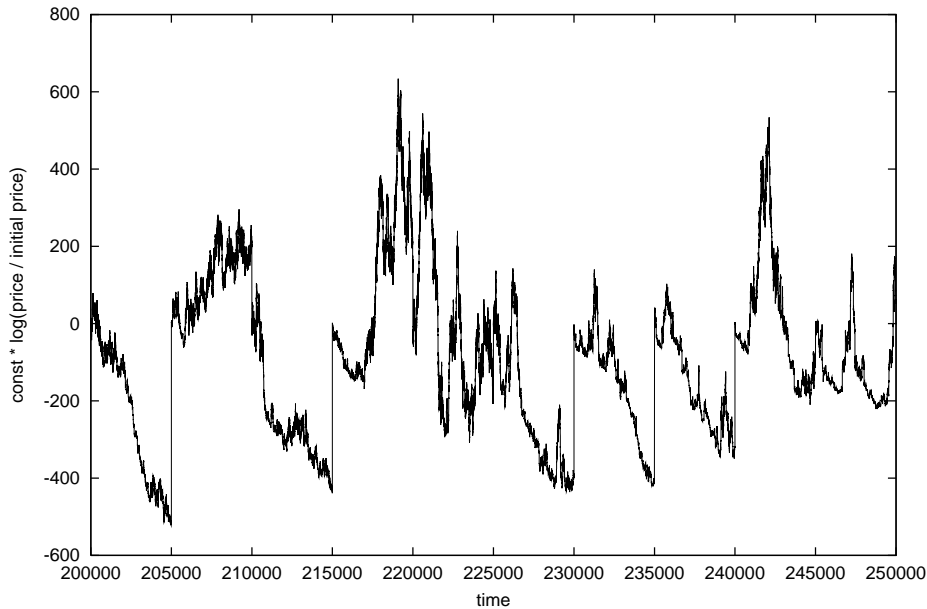}
\caption[]{Examples of price versus time in a modified Cont-Bouchaud model,
showing sharp peaks and flat valleys \cite{Changetal}}
\label{lux2}
\end{figure}

After this review, more effects were understood and variations were tried.
The crossover towards a Gaussian return distribution for increasing activity
was explained \cite{Kertesz}.
Instead of lattices of various dimensions, the Barabasi network was assumed
as the home of the Cont-Bouchaud traders, with good results \cite{Kullmann}.
Thermodynamic Ising-like modifications \cite{Silva}, in the direction of the
Iori model \cite{Iori} were proposed and gave reasonable
histograms of price fluctuations. The lack of time-reversal invariance was
recovered by putting some risk aversion psychology into the buying and selling
probabilities \cite{Chang}. Multifractality was found \cite{Castiglione} in
the higher moments of the return distributions for different time intervals.
Also a combination of these various modifications worked reasonably though
not ideally \cite{Changetal}; see e.g. Figure \ref{lux2}.

Applications included triple correlations between Tokyo, Frankfurt
and New York markets \cite{Schulze} and the effects of a small
Tobin tax on all transactions \cite{Ehrenstein,Westerhoff}.
Two physicists, Ehrenstein and Stauffer and an economist, Westerhoff
first independently simulated how such a tax would influence the market.
Depending on parameters, either the total tax revenue has a maximum in the
region of up to one percent taxation, or it increases monotonically.
Taking into account the tendency of governments to overexploit available
sources of tax income, they recommend the Tobin tax only for the first
case, not the second. It then would reduce the amount of speculation, but not
by an order of magnitude \cite{Ehrenstein/Westerhoff/Stauffer}. Summarizing, it
therefore appears that the Cont-Bouchaud models and the subsequent
variations on their theme contributed by other authors have gone
some way in explaining important stylized facts of financial
markets. Nevertheless, economists often feel somewhat uneasy about
this approach. The reason is that its starting point is known
knowledge about the characteristics of certain graph-based
dynamics (i.e., percolation models in statistical physics). The
``explanation'' of stylized facts in economics is, then, achieved
to some extend via a mere relabeling of known physical quantities
into new ones with an economic interpretation. Economists, of
course, would like to start with basic facets of economics
interaction of real-life markets rather than with a lattice-based
architecture. Furthermore, the many attempts at ``improvements''
of the model outlined above show that realistic results are only
obtained under extremely specific settings. Hence, it appears
questionable whether this framework really allows an explanation of
empirical findings that is ``independent of microscopic details''
as postulated in an econophysics manifesto (Stanley et al.
\cite{Stanley}).

\section{Social Percolation and Marketing: Solomon-Weisbuch}\label{SolomonWeisbuch}

Nevertheless, scientists have a natural tendency to apply what they have learned to as many
different problems as possible (maximizing thereby the number of their publications). Percolation
theory seems to be one of the discoveries one can use in quite a number of fields. Besides financial
markets, another application concerns ``social percolation'' and its use in marketing \cite{SolomonIII} which
we review in the following (departing shortly from our central subject of financial markets).

As in the previous section, every site of a large lattice is randomly either
occupied or empty, and a cluster is a set of occupied neighbors. Now we
identify occupied sites with potential customers of one specific product,
say, a Hollywood movie. Each site $i$ has a certain product quality requirement
$p_i$, and the product has a certain quality $q$. The values of $p_i$
are homogeneously distributed between zero and unity. Only those people visit
this movie (or more generally, buy this product) who regard its quality as
sufficient, i.e. who have $p_i < q$. We thus define as occupied a site with
$p_i < q$, and then a site is occupied with probability $q$. All sites $i$ in
a cluster have $p_i < q$.

If all potential customers are immediately informed about the new
product and its quality $q$, then a fraction $q$ of all sites will
buy, a trivial problem. But since we get so much advertising, we
may mistrust it and consider buying a movie ticket only if we hear
from a neighbor about its quality $q$. Thus a site $i$ buys if and
only if one of its nearest neighbors on the lattice has bought
before, if $i$ has not bought before, and if $p_i < q$ (customers
are assumed to have the same perception of the quality of the product, i.e. the quality assessment $q$ they tell
their neighbors is the same for all customers.) Initially, all
occupied sites on the top line of the lattice (top plane in three
dimensions) are regarded as customers who have bought and who thus
know the quality $q$. In this way, geometry plays a crucial role,
and only those sites belonging to clusters which touch the upper
line get informed about the movie and see it. If and only if one
cluster extends from top to bottom of the lattice, we say that the
cluster percolates or spans. And standard percolation theory then
predicts a spanning cluster if the occupation probability (or
quality) $q$ is larger than some threshold $p_c$, which is about
0.593 on the square lattice \cite{StaufferIII,Sahimi,Bunde}.
(Instead of starting from the top and moving towards the bottom,
one may also start with one site in the center and move outwards.
The cluster then percolates if it touches the lattice boundary.)

\begin{figure}
\centering
\includegraphics[angle=-90,width=0.9\textwidth]{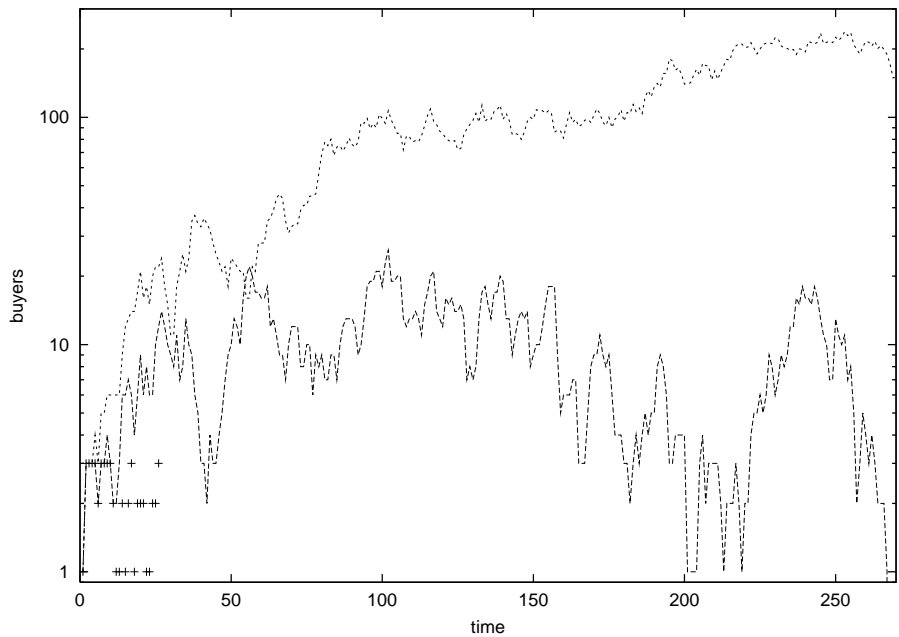}
\caption[]{Examples of social percolation starting with one central site
occupied; $q = p_c - 0.05$ (plus signs), $q = p_c$ (middle line), $q = p_c +
0.05$ (upper line). We plot logarithmically the number of new buyers in each
time interval \cite{Goldenberg}.
For the first four time steps the three buying curves agree,
since the same random numbers have been
used.}
\label{lux3}
\end{figure}

In this way the decades-old percolation theory divides new products into two
classes: Hits and flops \cite{Weisbuch}, depending on whether or not the
quality was high
enough to produce a spanning cluster. For a flop, $q < p_c$, only sites near the
initially occupied top line get informed that the new cluster exists, while
for a hit, $q > p_c$, also customers ``infinitely'' far away from the starting
line buy the movie ticket. In the case of a hit, except if $q$ is only very slightly
larger than $p_c$, nearly all customers with $p_i < q$ get the information and
go to the movie; only a few small clusters are then isolated from the spanning
network and have no chance to react. Figure \ref{lux3} shows three examples, where
one time step corresponds to one search through all neighbors of previously
occupied sites (Leath algorithm \cite{Stauffer_alt}).

In traditional marketing theories, as discussed in \cite{Goldenberg}, one
neglects the spatial structure, and a growing market has an exponential
time dependence until saturation sets in. For averages over lattices with
spanning clusters,
instead we have power laws in time \cite{Goldenberg}. In reality, both
cases have been observed, in addition to complicated behavior somewhat
similar to the curve $q = p_c$ of Figure \ref{lux2}.

The first modification of this static percolation model is to assume that
the quality $q$ changes in time: When a movie was successful (i.e. when
the cluster percolated), the producer lowers the quality of the next movie
slightly; when it was a flop (no percolating cluster), $q$ is slightly
increased. With this dynamics, like $q \rightarrow q \pm 0.001$, the $q$
automatically moves to the threshold $p_c$, an example of self-organized
criticality. In addition, we may assume \cite{SolomonIII} that also the $p_i$
change: $p_i$ increases by a small amount if $i$ just has seen a movie,
and it decreases by the same amount if the agent did not see a movie previously
(in the second case one has to distinguish whether the customer refused to
buy because of $p_i > q$ or merely was not informed about the movie.) In
this case also the $p_i$ can move towards $p_c$, though slower than $q$,
or they may be blocked at some intermediate value; also instabilities can
occur where $q$ and all $p_i$ move towards infinity. These difficulties were
clarified by Huang \cite{HuangI}, who also applied this model to stock markets
\cite{HuangII}.

Information through advertising influences the percolative phase
transition \cite{Proykova}. We refer to the literature cited above
as well as to \cite{Gupta,Ahmed,WeisbuchII,WeisbuchIII}\\\cite{WeisbuchIV,Sinha}
for further details and modifications.

There is (of course) also a large body of economic research
dealing with similar problems. In fact, the analysis of {\em
irreversible lock-in} and {\em path dependence} in the adaption of
new goods or new technologies is often based on mass-statistical
models. A prominent example is Arthur \cite{ArthurII} who used
nonlinear Polya urn models as an abstract model of such processes.
The application of similar ideas as an explanation for
geographical concentration of economic activity led to a
remarkable revival of the formerly dormant field of regional
economics over the nineties (cf. Arthur \cite{ArthurIII}, Krugman
\cite{Krugman}). Multi-agent approaches to ``hits'' and ``flops''
in the movie industry (using Bose-Einstein dynamics) with
empirical applications can be found in De Vany and Walls
\cite{DeVany} and De Vany and Lee \cite{DeVanyII}.

With the next (and last) model we come back to financial markets.

\section{Speculative Interaction and the Emergence of Scaling Laws: Lux-Marchesi}\label{LuxMarchesi}

The model of Lux and Marchesi \cite{LuxV} has its roots in earlier
attempts of economists at introducing heterogeneity into
stochastic models of speculative markets. Inspired by the analysis
of herd behavior in ant colonies \cite{Kirman} and earlier
applications of statistical mechanics to various problems in
sociology and political sciences (Weidlich and Haag
\cite{Weidlich,WeidlichII}), a stochastic model of trading in
financial markets has been developed in \cite{LuxII}. The basic
ingredient of this contribution was a kind of mean-field dynamics
for the opinion formation process among speculators together with
a phenomenological law for the ensuing price adjustment in the
presence of disequilibria. Using the Master equation formalism, it
could be shown that the model is capable of generating ``bubbles''
with over- or undervaluation of the asset as well as periodic
oscillations with repeated market crashes. A detailed analysis of
the dynamics of second moments (variances and co-variances) was
added in \cite{LuxIV} where the potential explanatory power of
multi-agent models for the typical time-variations of volatility
in financial markets was pointed out.

The group interactions in this model have been enriched in
\cite{LuxIII} by allowing agents to switch between a chartist and
fundamentalist strategy. This more complicated dynamics was shown
to give rise to chaotic patterns in mean values of the relevant
state variables (the number of agents in each group plus the
market price). Numerical analysis of simulated chaotic attractors
showed that they came along with leptokurtosis (fat tails) of
returns, hence providing a possible explanation of one of the
ubiquitous stylized facts of financial data.

Both microscopic simulations as well as more detailed quantitative
analyses of the resulting time series appeared in Lux and Marchesi
\cite{LuxV,LuxVI} and Chen {\em et al.} \cite{Chen}. The fact that
these key issues were approached quite lately in the development
of this market model to some extent reflects a broader trend in
the related literature: As already pointed out above, almost all
the early simulation models developed in economics had the initial
goal of investigating the formation of expectations of economic
agents in out-of-equilibrium situations (where it is hard to form
``rational'', i.e., correct expectations about the future) and
analyzing the selection of equilibria in the presence of multiple
consistent solutions of a static framework. Interestingly, a
development similar to that of the Lux-Marchesi model can also be
observed in the case of the Santa Fe Artificial Stock Market.
Although the latter was constructed by a group of researchers from
economics, physics, biology and computer science already in the
eighties, an analysis of the statistical properties of the
resulting time series only appeared recently (LeBaron {\em et
al.}, \cite{LeBaronIII}).

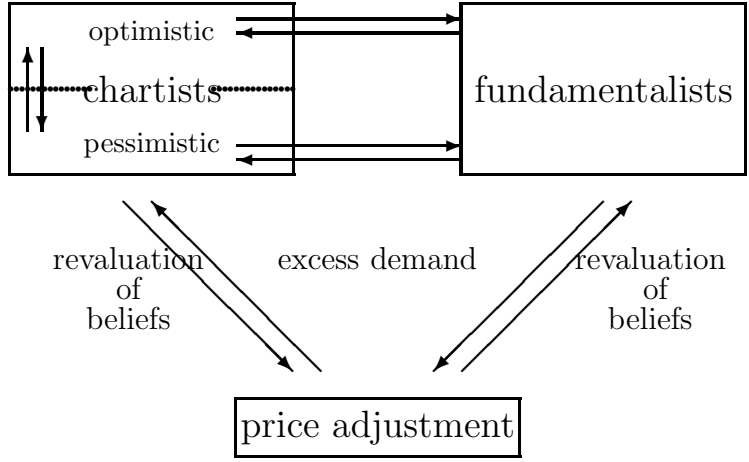
\begin{figure}[htb]
\setlength{\unitlength}{0.75cm}
\begin{picture}(16,9)
\thicklines
\put(0,0){\makebox(16,9){}}
\put(1.5,5,5){\framebox(5,3){\large{chartists}}}
\put(1.8,6.25){\vector(0,1){1.5}}
\put(2.05,7.75){\vector(0,-1){1.5}}

\put(9.5,5.5){\framebox(5,3){\large{fundamentalists}}}
\put(5.5,8.25){\vector(1,0){4}}
\put(9.5,8){\vector(-1,0){4}}
\put(5.5,6){\vector(1,0){4}}
\put(9.5,5.75){\vector(-1,0){4}}

\put(5.5,0.5){\framebox(5,1){\large{price adjustment}}}
\put(9.5,2){\vector(1,1){3}}
\put(12,5){\vector(-1,-1){3}}
\put(7,2){\vector(-1,1){3}}
\put(3.5,5){\vector(1,-1){3}}

\put(2.5,5.5){\makebox(3,1){\small{pessimistic}}}
\put(2.5,7.5){\makebox(3,1){\small{optimistic}}}
\multiput(1.5,7)(0.1,0){15}{\circle*{0.075}}
\multiput(6.5,7)(-0.1,0){15}{\circle*{0.075}}

\put(6.5,3.5){\makebox(3,1){\normalsize{excess demand}}}
\put(2,2.5){\makebox(2.75,1.5)}
\put(2.25,2,75){\shortstack{\normalsize{revaluation}\\\normalsize{of}\\\normalsize{beliefs}}}
\put(11.25,2.5){\makebox(2.75,1.5)}
\put(11.5,2.75){\shortstack{\normalsize{revaluation}\\\normalsize{ of}\\\normalsize{beliefs}}}
\end{picture}
\caption[]{Flowchart of dynamics of the Lux-Marchesi model: agents are allowed to switch between different behavioral alternatives. The number of individuals in these groups determines excess demand (the difference between demand and supply). Imbalances between demand and supply cause price adjustments which in turn affect agents' choices of strategies}
\label{lux_march}
\end{figure}

The dynamics of the ``terribly complicated'' (D.S.) Lux-Marchesi model is
illustrated in Figure \ref{lux_march}.

It is a kind of feedback between group
dynamics and price adjustment in the presence of imbalances
between demand and supply. Starting with basic definitions we denote by $N$ the total number of agents operating in our artificial market, $n_c$ the number of noise traders, $n_f$ the number of fundamentalists $(n_c+n_f=N)$, $n_+$ the number of optimistic noise traders, $n_-$ the number of pessimistic noise traders $(n_++n_-=n_c)$; $p$ is the market price, $p_f$ the fundamental value.

The dynamics of the model are composed of the following elements:

1. \textit{noise traders' changes of opinion} from a pessimistic
to an optimistic mood and \textit{vice versa}: the probabilities
for these changes during a small time increment $\Delta t$ are
given by $\pi_{+-} \Delta t$ and $\pi_{-+} \Delta t$ and are
concretised as follows:

\begin{eqnarray}
\pi_{+-} &=&  v_1\frac{n_c}{N}\exp(U_1)\nonumber\\
\pi_{-+} &=& v_1\frac{n_c}{N}\exp(-U_1) \nonumber\\
\ U_1 & = &
\alpha_1x+\frac{\alpha_2}{v_1}\frac{dp}{dt}\frac{1}{p}
\end{eqnarray}

(We denote by $\pi_{ab}$ the rates from state $a$ to state $b$, like
$a$ for optimism.)
Here, the basic influences acting on the chartists' formation of
opinion are the majority opinion of their fellow traders,
$x=\frac{n_+-n_-}{n_c}$, and the actual price trend,
$\frac{dp}{dt} \frac{1}{p}$. Parameters $v_1$, $\alpha_1$, and
$\alpha_2$ are measures of the frequency of revaluation of opinion
and the importance of ``flows'' (i.e. the observed behaviour of
others) and charts, respectively. Furthermore, the change in asset
prices has to be divided by the parameter $v_1$ for the frequency
of agents' revision of expectations since for a consistent
formalization one has to consider the $mean$ price change over the
$average$ interval between successive revisions of opinion. The
transition probabilities are multiplied by the actual fraction of
chartists (that means, it is restricted to such a fraction)
because chartists are also allowed to interact with fundamental
traders in the second component of the group dynamics that follows
below.

2. \textit{switches between the noise trader and fundamentalist
group} are formalised in a similar manner. Formally, one has to
define four transition probabilities, where the notational
convention is again that the first index gives the subgroup to
which a trader moves who had changed her mind and the second index
gives the subgroup to which she formerly belonged (hence, as an
example, $\pi_{+f}$ gives the probability for a fundamentalist to
switch to the optimistic chartists' group):

\begin{eqnarray}
\pi_{+f}=v_2\frac{n_+}{N}\exp(U_{2,1})\mbox{,} \quad \pi_{f+}=v_2\frac{n_f}{N}\exp(-U_{2,1}) \\
\pi_{-f}=v_2\frac{n_-}{N}\exp(U_{2,2})\mbox{,} \quad
\pi_{f-}=v_2\frac{n_f}{N}\exp(-U_{2,2}) \mbox{.}
\end{eqnarray}

The forcing terms $U_{2,1}$ and $U_{2,2}$ for these transitions depend on the difference between the (momentary) profits earned by using a chartist or fundamentalist strategy:

\begin{equation}
U_{2,1}=\alpha_3 \left \{ \frac{r+\frac{1}{v_2} \frac{dp}{dt}}{p}-R
- {s\cdot \Big |\frac{p_f-p}{p} \Big |} \right \} \
\end{equation}

\begin{equation}
U_{2,2}=\alpha_3 \left \{ \ { R- \frac{r+\frac{1}{v_2}
\frac{dp}{dt}}{p} }-{s\cdot \Big |\frac{p_f-p}{p} \Big |} \right \}\
\end{equation}

The first term of the U functions represents the profit of
chartists from the $n_+$ group and $n_-$ group. The second term is
the profit of the fundamentalists. The parameters $v_2$ and
$\alpha_3$ are reaction coefficients for the frequency with which
agents reconsider appropriateness of their trading strategy, and
for their sensitivity to profit differentials, respectively.
Excess profits (compared to alternative investments) enjoyed by
chartists from the optimistic group are composed of nominal
dividends ($r$) and capital gains due to the price change
($dp/dt$). Dividing by the actual market price gives the revenue
per unit of the asset. Excess returns compared to other investment
opportunities are computed by substracting the average real return
($R$) received by the holders of other assets in our economy.
Fundamentalists, on the other hand, consider the deviation between
price and fundamental value $p_f$ (irrespective of its sign) as
the source of arbitrage opportunities from which they may profit
after a return of the price to the underlying fundamental value, so that
a large difference between $p$ and $p_t$ induces traders to follow the fundamentalist strategy.
As the gains of chartists are immediately realised whereas those
claimed by fundamentalists occur only in the future (and depend on
the uncertain time for reversal to the fundamental value) the
latter are discounted by a factor $s<1$. Furthermore, neglecting
the dividend term in fundamentalists' profits is justified by
assuming that they correctly perceive the (long-term) real returns
to equal the average return of the economy (i.e. $r/p_f=R$) so
that the only source of excess profits in their view is arbitrage
when prices are ``wrong'' ($p \not= p_f$). As concerns the second
U-function, $U_{2,2}$ one considers profits from the viewpoint of
pessimistic chartists who in order to avoid losses will rush out
of the market and sell the asset under question. Their fall-back
position by acquiring other assets is given by the average profit
rate $R$ which they compare with nominal dividends plus price
change (which, when negative, amounts to a capital $loss$) of the
asset they sell. This explains why the first two items in the
forcing term are interchanged when proceeding from $U_{2,1}$ to
$U_{2,2}$.

3. \textit{price changes} are modelled as endogenous responses of
the market to imbalances between demand and supply. Assuming that
optimistic (pessimistic) chartists enter on the demand (supply)
side of the market, excess demand (the difference between demand
and supply) of this group is:

\begin{equation}
ED_c=(n_+-n_-)t_c
\end{equation}
with $t_c$ being the average trading volume per transaction.
Fundamentalists' sensitivity to deviations between market price
and fundamental value leads to a law of the type:

\begin{equation}
ED_f=n_f \cdot \gamma \frac{p_f-p}{p}
\end{equation}
with $\gamma$ being a parameter for the strength of reaction. In
order to conform with the general structure of this framework,
the price adjustment process is also formalised in terms of (Poisson)
transition probabilities. In particular, the transition probabilities
for the price to increase or decrease by
a small percentage $\Delta p=\pm 0.001 p$ during a time increment
$\Delta t$ are given by:\footnote{The increment $\Delta p$ has been chosen as
small as possible in order to avoid artificial lumpiness of price
changes with concentration of the distribution of returns at a few
values only.}

\begin{eqnarray}
\pi_{\uparrow p}=\max[0,\beta (ED+\mu )] &\mbox{,}& \pi_{\downarrow
p}=-\min[\beta (ED+\mu ),0]\end{eqnarray} where $\beta$ is a
parameter for the price adjustment speed and $ED=ED_c+ED_f$ is
overall excess demand (the sum of excess demand by both noise
traders and fundamentalists).

This probabilistic rule for price adjustments is, in fact, equivalent to the traditional Walrasian adjustment scheme. It can be shown that the mean value dynamics of the price can be depicted by the textbook differential equation for the dependence of price changes on overall excess demand:

\begin{equation}
\frac{dp/dt}{p}=\beta\cdot ED=\beta\cdot(ED_c+ED_f)
\end{equation}

Note that these price changes feed back on agents' decisions to
follow one or the other trading strategy: a price increase will
reinforce optimistic beliefs and will make formerly pessimistic
chartists join a bullish majority. Similarly, price changes might
bring $p$ closer to an assumed fundamental value, $p_f$, which
strengthens fundamentalist beliefs, or they might lead to larger
deviations from $p_f$  which reinforces the position of chartists.
All in all, the resulting confirmation or disappointment of
agents' opinions together with changing profitability of
strategies will lead to switches between groups altering the
composition of the population and effecting excess demand of the
following period. The model also allows for exogeneous changes of
the fundamental value:

4. \textit{changes of fundamental value}: in order to assure that
none of the stylised facts of financial prices can be traced back
to exogenous factors, it is assumed that the log-changes of $p_f$ are
Gaussian random variables:
$\ln(p_{f,t})-\ln(p_{f,t-1})=\varepsilon_t$ and $\varepsilon_t
\sim N(0,\sigma_{\varepsilon})$. The Poisson type dynamics of
asynchronous updating of strategies and opinions by the agents can
only be approximated in simulations. In particular, one has to
choose appropriately small time increments in order to avoid
artificial synchronicity of decisions. In [35, 108, 109] a
simulation program with some flexibility in the choice of the time
increment is used. Namely, time increments $\Delta t=0.01$ are
used for ``normal times'', while during volatility bursts the
precision of the simulations was automatically increased by a
factor 5 (switching to $\Delta t=0.002$) when the frequency of
price changes became higher than average. This procedure requires
that all the above Poisson rates be divided by 100 or 500,
(depending on the precision of the simulation) in order to arrive
at the probability for any single individual to change his
behaviour during [$t$, $t+\Delta t$]. Similarly, it is assumed
that the auctioneer adjusts the prevailing price by one elementary
unit (one cent or one pence) with probabilities $w_{\uparrow p}$
or $w_{\downarrow p}$ during one time increment. For the time
derivative, $dp/dt$, the average of the prices changes during the
interval [$t-10\Delta t$, $t$] has been used. Furthermore,
occurence of the ``absorbing states'' $n_c=0$ ($n_f=N$) and
$n_c=N$ ($n_f=0$) was excluded by setting a lower bound to the
number of individuals in both the group of chartists and
fundamentalists.

\begin{figure}
\centering
\includegraphics[angle=270, width=1.1\textwidth]{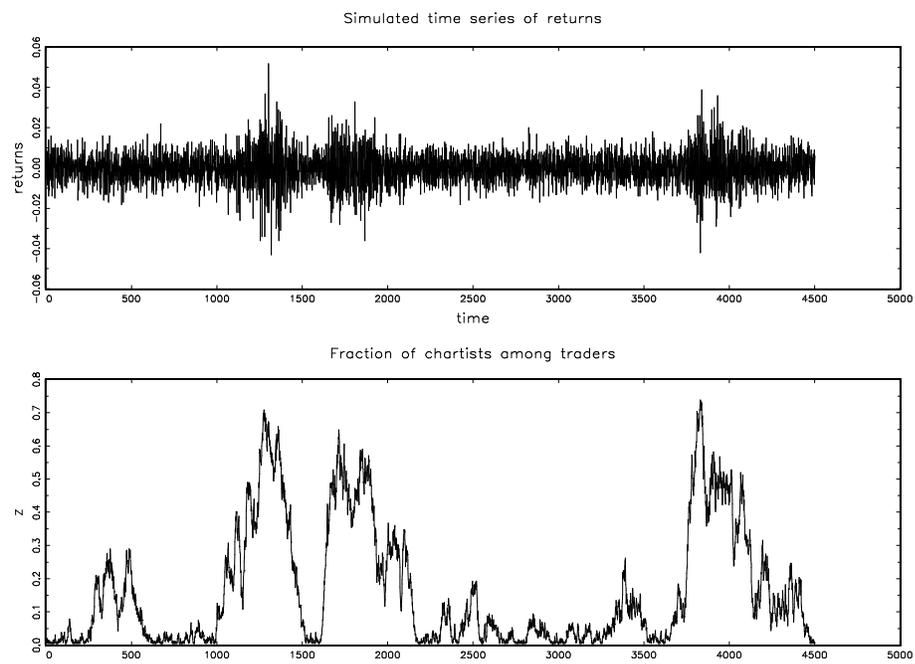}
\caption[]{Time series of returns (relative price changes, upper panel) and the fraction of chartists (lower panel) from a typical simulation of the Lux-Marchesi model}
\label{review}
\end{figure}

The overall results of this dynamics is easily understood by
investigation of the properties of stationary states (cf.
\cite{LuxVI}), i.e., situations in which there are no predominant
flows to one of the groups and the price remains constant. Such a
scenario requires that there is a balanced disposition among
(chartist) traders, i.e., we neither have a dominance of optimists
over pessimists (nor {\em vice versa}) and that the price is equal to the
fundamental value (which makes fundamentalists inactive). A little
reflection reveals that in such a situation, there is no advantage
to either the chartist or fundamentalist strategy: no misprizing
of the asset nor any discernible trends exist. Hence, the
composition of the population becomes {\em indeterminate} which
implies that, in the vicinity of these stationary states, the
group dynamics is governed only by stochastic factors. \footnote{A similar indeterminacy
in the number of agents in different groups has been found in a model of resource
extraction (Youssefmir and Huberman \cite{Youssefmir}). They also emphasize that
this indeterminacy can lead to burst of activity (temporary large fluctuations). Another
recent example of similar intermittent dynamics appears in an artificial foreign
exchange market in which agents are using genetic algorithms to adopt their portfolio strategy
to changing circumstances, cf. \cite{ArifovicGen} \cite{LuxundSchorn}.} Hence, to a
first approximation one can abstract from the economic forces
which apparently become relevant only in out-of-equilibrium situations. As
detailed in \cite{LuxV,LuxVI}, the stationary states
described above may either be locally stable or unstable with the
number of chartists acting as a bifurcation parameter. Simulations
show that temporary deviations into  the unstable region can be
interpreted as intermittent behavior which generates clusters of
volatility and numerically accurate power laws for the tail
behavior of raw returns as well as long-term dependence  in absolute
and squared returns. Figure \ref{review} illustrates the
interplay between the dynamics of relative price changes and the
development of the number of chartists among traders.

As can be directly inferred from the graph, an increase of the
number of chartists leads to intermittent fluctuations. Note also
that the model incorporates self-stabilizing forces leading to a
reduction of the number of chartists after a period of severe
fluctuations. The reason is that large deviations of the price
from its fundamental value lead to high potential profits of the
fundamentalist strategy which induces a certain number of agents
to switch away from chartism. Chen {\em et al.} \cite{Chen} also
show that the motion of the market price appears totally random
(judged by standard tests for determinism and nonlinearity) in
tranquil times but shows traces of non-linear structure during
more volatile episodes \cite{Chen}. This feature appears to be in
harmony with findings for the U.S. stock market
\cite{Lima}. Recent work in this area has come up with some rigorous
results on the statistical properties of simpler variants of this type
of models and has used these characterizations in order to estimate
the parameters governing agent's interactions, cf.
\cite{AlfaranoI} \cite{AlfaranoII} \cite{AlfaranoIII}.

\section{Discussion}

While early attempts at microscopic simulations of financial
markets appeared unable to account for the ubiquitous scaling laws
of returns (and were, in fact, not devised to explain them), some
of the recent models seem to be able to explain some of the
statistical properties of financial data (usually denoted as
``anomalies'' in economics). Nevertheless, there is still a number
of important topics left to future research: first, the recent
surge of newly available data on  the intra-daily level has opened
a Pandora's box of new regularities on very small time scales (cf.
 Dunis and Zhou \cite{Dunis}). While the
ubiquitous scaling laws found in all markets might be explained
well by simple mechanisms beloved by physicists, the more delicate
intra-daily patterns may require more detailed models (denoted as
``monsters'' in a workshop presentation of a paper by
Maslov \cite{Maslov}). If physicists do not want to stop half-way
in their contribution to economics, they may probably have to
develop, as is typically done in economics, models with more
institutional background \cite{Madhavan}.
\footnote{Similar probably to the
development of statistical models of traffic flows, cf. Nagel {\em
et al.} \cite{Nagel}} Second, although we have a bunch of models
for power-laws, their generality is still restricted in one very
important respect: the ``interesting'' dynamics does only apply
for a certain range of the population size $N$ of speculators and
in most cases does not survive for realistically large $N$. This has been
shown for the Kim and Markowitz and Lux and Marchesi models in
Egenter {\em et al.} \cite{Egenter} and probably applies to most
alternative approaches. A recent investigation of Chen and Yeh's
artificial stock market also shows that their interesting results
tend to vanish when the number of traders increases (cf. Yeh
\cite{Yeh}). In Lux and Marchesi, the finite size effect
immediately becomes apparent by realizing that the overall number
of agents affects excess demand and, therefore, the right-hand
side of the price-adjustment equation. However, although one might
expect that this leads to more severe fluctuations with increasing
$N$, the contrary is the case: fluctuations become dampened with
higher $N$  and finally die out altogether with a cross-over of
returns to a Normal distribution. Of course, linear dependence of
excess demand on $N$  is not realistic. The task for future
research is, therefore, to look for self-organizing forces in the
market (maybe via the addition of wealth dynamics) which may lead
to an effective confinement of the level of excess demand.

Have the econophysics papers reviewed here brought anything new to
economics? Certainly they did not invent microscopic and agent-based
computer modeling (http://www.complexity-research.org/mad)
of markets or empirical analysis of market fluctuations.
But the large number of enonophysicists pushed these areas since
physicists are more familiar with computer simulation and
empirical analysis than many mainstream economists more interested
in exact mathematical solutions. Of course, percolation and random
field Ising models are clear physics contributions, and the
introduction of multi-fractal processes as stochastic models of
financial prices (a topic which is outside the scope of the present review)
is conceived as an important innovation by many
economists. Here again, we find that economists have been aware of
the multi-scaling of returns for some time \cite{Ding,Lux}, but suffered from a
lack of appropriate models in their tool-box (cf. \cite{LuxVII}
for more details on this issue).

Have econophysicists made any predictions which were later
confirmed? If we define as ``prediction'' something which has
appeared in a journal on paper before the predicted event was
over, we exclude all private communications or e-prints, and know
only three cases: The warning published in September 1997 that a
``krach'' should occur before the end of November \cite{Dupois}
(it occurred in October); the assertion that the Nikkei index
in Tokyo should go up in 1999, which it did by roughly the
predicted amount, and the prediction that the U.S. stock market should
reach a lower turning point in early 2004 which did not happen \cite{SornetteX}.
Even if ``successful'', relatively vague predictions like the above are, of
course, at best interpreted as anecdotical evidence, but are surely not
significant from a statistical perspective.

Have we become rich in this way? The senile co-author gained 50\%
in half a year by believing the above predictions, and similar
anecdotal evidence exists from others. Interestingly, in this way
the one contributer with a physics background seems to show a
better performance in private portfolio management than the three
economists who rather concentrated on their academic and professional careers. Of
course success is often reported proudly while failures are kept
as a secret. In this way, certain strategies might appear
successful simply because of a bias in awareness of positive
outcomes versus negative ones. More than half of a century ago,
the prominent economist Nicholas Kaldor \cite{Kaldor} explained
the prevalence of chartist strategies by such a misperception of
their track record. But more reliable are the flourishing
companies like Prediction Company (New Mexico) or Science-Finance
(France) founded by physicists Farmer and Bouchaud, respectively,
together with economists, giving employment to $10^2$ people. This
seems quite a success for theoretical physicists.
\newline
\newline
\newline
\newline
\newline
\newline
{\large {\bf Acknowledgment}}
\newline
\newline

The authors are extremely grateful to Florian Heitger for his help with the
preparation of a Latex file of this review.

\bibliographystyle{plain}
\bibliography{Lit_article,Lit_thesis,Lit_book,Lit_bookedit,Lit_manuscript}
\end{document}